\documentclass[fleqn,usenatbib]{mnras}

\usepackage{newtxtext,newtxmath}

\usepackage[T1]{fontenc}

\DeclareRobustCommand{\VAN}[3]{#2}
\let\VANthebibliography\thebibliography
\def\thebibliography{\DeclareRobustCommand{\VAN}[3]{##3}\VANthebibliography}

\makeatletter
\def\@fnsymbol#1{\ensuremath{\ifcase#1\or *\or   \dagger\or \ddagger\or
   \mathsection\or \mathparagraph\or \|\or **\or \dagger\dagger
   \or \ddagger\ddagger \else\@ctrerr\fi}}
\makeatother

\newcommand{\flamingo}{\texttt{FLAMINGO}}
\newcommand{\simba}{\texttt{SIMBA}}
\newcommand{\antilles}{\texttt{ANTILLES}}
\newcommand{\bahamas}{\texttt{BAHAMAS}}
\newcommand{\fable}{\texttt{FABLE}}
\newcommand{\xfable}{\texttt{XFABLE}}
\newcommand{\eagle}{\texttt{EAGLE}}

\newcommand{\Masa}[1]{{\bfseries\textcolor{blue}{[Masa: #1]}}}
\newcommand{\Leah}[1]{{\bfseries\textcolor{purple}{[Leah: #1]}}}
\newcommand{\amon}[1]{{\bfseries\textcolor{pink}{[Alex: #1]}}}

\usepackage{amsmath}	
\usepackage{newtxtext,newtxmath}
\usepackage{pdfpages}
\usepackage{float}
\usepackage{tabularx}
\usepackage{comment}
\usepackage[normalem]{ulem}
\usepackage[mathlines, pagewise]{lineno}
\usepackage{colortbl}

\usepackage{subcaption}
\usepackage{graphicx}	
\usepackage{amsmath}	
\usepackage{hyperref}
\usepackage{physics}
\usepackage{arydshln}
\usepackage{multirow}

\title[The kSZ effect in simulations]{The kinetic Sunyaev Zeldovich effect as a benchmark for AGN feedback models in hydrodynamical simulations: insights from DESI + ACT}

\author[L. Bigwood et al.]{
\parbox{\textwidth}{
\Large Leah Bigwood,$^{1,2}$\thanks{E-mail: lmb224@cam.ac.uk}
Masaya Yamamoto,$^{3}$ 
Jared Siegel,$^{3}$
Alexandra Amon,$^{3}$
Ian G. McCarthy,$^{4}$ \\
Romeel Dave,$^{5,6,7}$
Jaime Salcido,$^{4}$
Matthieu Schaller,$^{8,9}$
Joop Schaye,$^{8}$
Tianyi Yang,$^{4}$\\
\small
$^{1}$Institute of Astronomy, University of Cambridge, Madingley Road, Cambridge, CB3 0HA, UK\\
$^{2}$Kavli Institute for Cosmology (KICC), University of Cambridge, Madingley Road, Cambridge CB3 0HA, UK\\
$^{3}$Department of Astrophysical Sciences, Princeton University, 4 Ivy Lane, Princeton, NJ 08544, USA\\
$^{4}$Astrophysics Research Institute, Liverpool John Moores University, Liverpool, L3 5RF, UK\\
$^{5}$Institute for Astronomy, Royal Observatory, Univ. of Edinburgh, Edinburgh EH9 3HJ, UK\\
$^{6}$University of the Western Cape, Bellville, Cape Town 7535, South Africa\\
$^{7}$South African Astronomical Observatories, Observatory, Cape Town 7925, South Africa\\
$^{8}$Leiden Observatory, Leiden University, PO Box 9513, 2300 RA Leiden, the Netherlands\\
$^{9}$Lorentz Institute for Theoretical Physics, Leiden University, PO box 9506, 2300 RA Leiden, the Netherlands}}

\date{Accepted XXX. Received YYY; in original form ZZZ}

\pubyear{\the\year{}}

\begin{document}
\label{firstpage}
\pagerange{\pageref{firstpage}--\pageref{lastpage}}
\maketitle

\begin{abstract}
Baryonic feedback remains one of the largest uncertainties in cosmological hydrodynamical simulations, with different prescriptions producing divergent predictions for the fraction of gas expelled from halos, the radial extent of the gas expulsion and the impact on large scale matter clustering.  We present the first systematic study of the kinetic Sunyaev–Zel’dovich (kSZ) effect across a wide range of  simulations (FLAMINGO, ANTILLES, BAHAMAS, SIMBA, FABLE and their variants),  and compare them directly to DESI Year 1 + ACT kSZ measurements.
We ensure a like-for-like comparison with observations by developing a robust methodology that accounts for the halo mass selection using galaxy-galaxy lensing, cosmic variance, miscentering and satellites, establishing the kSZ effect as a new benchmark for the simulations. 
We find that fiducial feedback models are disfavoured by $>3\sigma$, while simulations with more powerful AGN feedback within the FLAMINGO and BAHAMAS suites, as well as SIMBA, reproduce the observed kSZ signal within $<2\sigma$.
We use the ANTILLES simulation suite to demonstrate that the amplitude of the kSZ effect is a strong predictor of matter power spectrum suppression, competitive with baryon fraction metrics. These results establish the kSZ as a critical probe for evaluating feedback physics and for advancing the fidelity of cosmological simulations.

\end{abstract}




\section{Introduction}

Baryon feedback redistributes gas within halos and into the intergalactic medium, impacting the large-scale matter distribution, but the mechanism for doing this is not well understood.
Cosmological hydrodynamical simulations can be made to reproduce a wide range of galaxy, group, and cluster properties, yet they diverge substantially in their predictions for the fraction of gas expelled, the radial extent of this expulsion, and its imprint on matter clustering \citep[e.g.][]{McCarthy2017,Schaye2023,Pakmor2023,Bigwood2025,Dave2019, vandaalen:2020, Semboloni2011}.  These differences stem from uncertain prescriptions for stellar and AGN feedback, which regulate star formation and the thermal state of halo gas, but whose efficiency and coupling remain poorly constrained by galaxy data alone. Because feedback occurs below the resolution scale of simulations, empirical subgrid models are employed, typically calibrated to the galaxy stellar mass function (GSMF) and/or X-ray gas fractions in clusters. 
Progress requires benchmarks beyond the traditional observables to which feedback models have been tuned.

In this work, we lay out a roadmap for robustly studying the kinetic Sunyaev–Zel’dovich (kSZ) across a range of hydrodynamical simulations, and for establishing direct comparison with observations. Measurements of the kSZ effect \citep{SZ1972,SZ1980} have now come to fruition, enabled by high-resolution cosmic microwave background (CMB) experiments such as the Atacama Cosmology Telescope \citep[ACT,][]{act2016,act2020} and wide-area, highly multiplexed spectroscopic surveys such as the Dark Energy Spectroscopic Instrument \citep[DESI,][]{desi2022}, which together allow the stacking of its relatively weak imprint weighted by halo velocity. After decades of groundwork, stacked kSZ measurements have now traced radial gas profiles \citep{schaan2016,battaglia2017,Schaan2021,mallaby2023,Hadzhiyska2024,Ried2025, Roper2025}. Compared to X-ray gas fraction data, kSZ measurements probe the larger-scale radial distribution of gas, and do so for halos in a complementary mass–redshift regime, owing to their direct proportionality to electron number density \citep[e.g.][]{Battaglia_SNOWMASS}. The kSZ effect provides a new opportunity to differentiate feedback models, crucial for breaking degeneracies that limit consensus on the impact of feedback on the matter distribution and for uncovering the underlying physical mechanisms. We focus on the most recent spectroscopic measurements from DESI cross-matched with ACT CMB maps \citep{Ried2025}.

Early kSZ analyses have indicated a preference for a `strong' feedback scenario; either by constraining a larger suppression of the matter power spectrum than predicted by many state-of-the-art hydrodynamical simulations \citep{Schneider2022, Bigwood2024}, or by demonstrating that the observed signal is incompatible with simulations displaying too weak feedback \citep{McCarthy2024,Hadzhiyska2024,Ried2025, Hadzhiyska2025, kovac2025,siegel2025b}. Independent evidence has since accumulated from complementary probes: recent measurements of the thermal Sunyaev–Zel’dovich (tSZ) Compton Y - halo mass relation \citep{Dalal25}, tSZ and cosmic shear cross-correlation \citep{Pandey25}, the diffuse X-ray background \citep{Ferreira2024,laposta2024} and fast radio bursts \citep{Reischke2025}; all corroborating the picture of enhanced matter power suppression by baryonic feedback.  In addition, new X-ray measurements from eROSITA suggest that galaxy groups are more gas-depleted \citep{Popesso2024, Dev2024, siegel2025b}, and therefore subject to stronger AGN-driven gas expulsion, than indicated by previous X-ray observations which many simulations and models are calibrated to (see Figure~5 of \citealt{siegel2025b}).  Taken together, these results highlight growing tension between simulations and observations \citep[see, e.g.][]{salcido2025}, underscoring the need for a coherent feedback model that can reconcile all available constraints.

In this work we present the first systematic study of the kSZ effect across a broad suite of cosmological hydrodynamical simulations: \antilles\ \citep{Salcido2023}, \bahamas\ \citep{McCarthy2017}, \simba\ \citep{Dave2019}, \fable\ \citep{Henden2018}, and \flamingo\ \citep{Schaye2023}, as well as strong-feedback variants and the \antilles\ parameter hypercube.
We develop a robust methodology for simulation–data comparison, accounting for cosmic variance, halo mass calibration, mis-centering, and satellites, and confront simulations with the latest DESI+ACT kSZ measurements. We build upon the methodology established in \citet{McCarthy2024} and \citet{siegel2025b} to use galaxy-galaxy lensing (GGL) measurements of the projected surface mass density to select a simulated galaxy sample with consistent halo mass and satellite fraction, enabling a like-for-like comparison. This enables a decisive test of feedback models and their imprint on both halo gas and the large-scale matter distribution.


This paper is structured as follows. Section~\ref{sec:data} introduces the simulated and observational datasets, and Section~\ref{sec:methods} establishes the methodologies for measuring the kSZ effect in simulations. In Section~\ref{sec:measurement}, we examine the simulated kSZ signal under different galaxy selections and compare it to DESI+ACT observations. Section~\ref{sec:kszsummary} then presents the correlation between the kSZ signal and the matter power spectrum suppression and discusses the role of kSZ as a benchmark for hydrodynamical simulations.

\section{Data}\label{sec:data}

\subsection{Cosmological hydrodynamical simulations}\label{subsec:sims}
\begin{figure*}
\centering
\includegraphics[width=1\linewidth]{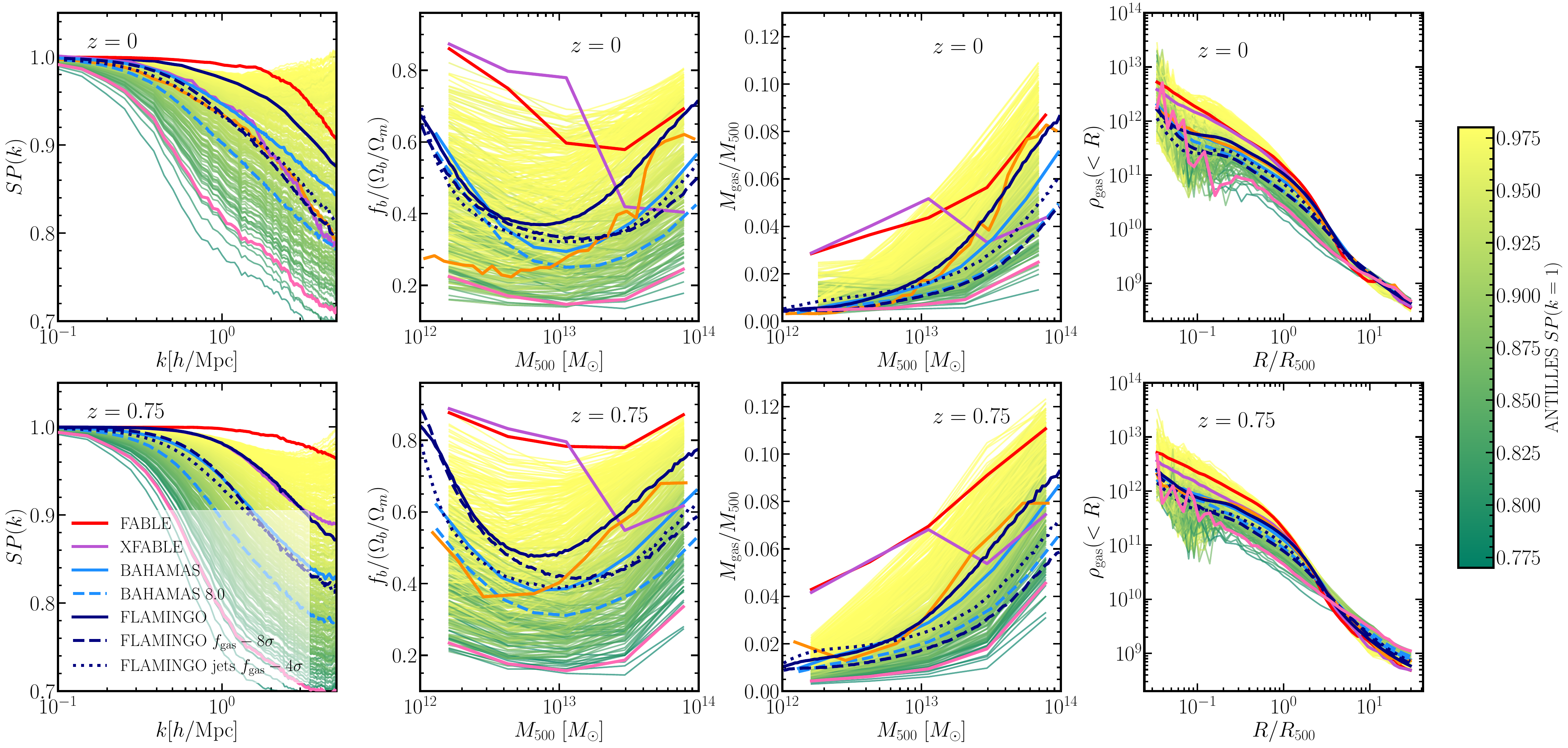}

\caption{\textit{Left: } The matter power spectrum suppression due to baryonic effects, $SP(k)=P(k)/P_{\mathrm{DM only}}(k)$, at $z=0$ (\textit{top}) and $z=0.75$ (\textit{bottom}), measured in each of the hydrodynamical simulations we study. Note that we do not measure $SP(k)$ at $z=0.75$ for \simba\, since the corresponding dark matter only simulation output is not available. 
\textit{Centre left: }The median total baryon fraction normalised by the universal baryon fraction, $f_{\rm b}/(\Omega_{\rm b}/\Omega_{\rm m})$, as a function of halo mass, $M_{500}$, measured in each of the simulations we study at $z=0$ (\textit{top}) and $z=0.75$ (\textit{bottom}). 
\textit{Centre right: }The median hot gas mass fraction, $M_{\mathrm{gas}}/M_{500}$, as a function of halo mass, $M_{500}$, measured in each of the simulations we study at $z=0$ (\textit{top}) and $z=0.75$ (\textit{bottom}).  We measure $M_{\mathrm{gas}}$ within a spherical aperture of radius $R_{500}$ centered on the particle with the minimum gravitational potential energy in the halo.  \textit{Right:} The radial hot gas density profile, showing the gas density enclosed within a radius $R$, plotted as a function of the radius normalized to $R_{500}$.  We plot the mean profile computed for halos with $13.25<\log_{10}(M_{500} [\mathrm{M_{\odot}}])<13.35$. In all panels we plot the 400 \antilles\ boxes, as well as \fable\ (red), \xfable\ (purple), \flamingo\ (navy solid), \flamingo$f_{\mathrm{gas}}-8\sigma$ (navy dashed), \flamingo\ jets $f_{\mathrm{gas}}-4\sigma$ (navy dashed), \bahamas\ (blue solid), \bahamas\ 8.0 (blue dashed) and \simba\ (orange).  The \antilles\ simulations are colour-coded by the suppression of the matter power spectrum due to baryonic effects measured at $k=1~\mathrm{Mpc}/h$. The pink line represents the \antilles\ box that provides the best combined fit to the DESI + ACT BGS, LRG M1 and LRG M2 kSZ measurements (see Section~\ref{sec:antillescompare}).}
\label{fig:simprops}
\end{figure*}

In the following sections, we describe the key properties and subgrid baryonic feedback models for the state-of-the-art cosmological hydrodynamical simulation suites we study: \bahamas, \flamingo, \simba, \fable\ (and \xfable) and \antilles.  Each simulation's basic properties are summarized in Table.~\ref{tab:sims}.  Furthermore, in Figure~\ref{fig:simprops} we plot the suppression of the matter power spectrum due to baryonic effects, the total baryon fraction as a function of halo mass, the gas mass fraction as a function of halo mass, and the radial gas density profile, measured in each of the simulations studied at both $z=0$ and $z=0.75$, with the latter being the relevant redshift for the DESI+ACT LRG samples.  We will refer to these plots to interpret and contextualize the measured kSZ signal from simulations throughout the work.  We note that \citet{Schaller2024} demonstrated that in FLAMINGO-like simulations, the suppression of the matter power spectrum was only converged for boxes with volume larger than $(200 ~\mathrm{Mpc}/h)^3$, meaning it is possible that this measurement is unreliable in the smaller simulation boxes we study.

\begin{table*}
    \centering
    \caption{\label{tab:sims} The key properties of the simulation boxes we study.  We list the box size, the mass of baryonic particles, $m_{\rm gas}$, the AGN feedback modes implemented, the $z=0$ suppression of the matter power spectrum  $SP(k)$ measured at $k=1h/\mathrm{Mpc}$, the $z=0$ median baryon fraction $f_{\rm b}/(\Omega_{\rm b}/\Omega_{\rm m})$ measured at $M_{500}=10^{13.25}M_\odot$, and the cosmology adopted.}

    \begin{tabular}[width=\textwidth]{cccccccccc}
        \hline\hline
             Simulations & Volume [$\mathrm{Mpc}^3$] & $m_{\rm gas}$ [$\mathrm{M_{\odot}}$] & AGN feedback mode & $\frac{f_{\rm b}}{(\Omega_{\rm b}/\Omega_{\rm m}) }(10^{13.25}M_\odot)$ &  $SP(1h/\mathrm{Mpc})$  & Cosmology \\ \hline
             ANTILLES & $(100 ~h^{-1})^3$ & $1.09 \times 10^{9}$ & Thermal& 0.142 - 0.690 & 0.776 - 0.983 & WMAP 9-year\\
             BAHAMAS & $(400 ~~h^{-1})^3$ & $1.14 \times 10^{9}$ & Thermal& 0.316 &  0.903 &  WMAP 9-year \\
             BAHAMAS 8.0 & $(400 ~~h^{-1})^3$ & $1.14 \times 10^{9}$ & Thermal& 0.255 & 0.947 & WMAP 9-year \\
             FLAMINGO & $(1000)^3$ & $1.0 \times 10^{9}$ & Thermal& 0.420 & 0.976 & DES Y3 3$\times$2pt\\
             FLAMINGO $f_{\rm gas}-8\sigma$ & $(1000)^3$ & $1.0 \times 10^{9}$ & Thermal& 0.322 & 0.333  & DES Y3 3$\times$2pt\\
             FLAMINGO jets $f_{\rm gas}-4\sigma$ & $(1000)^3$ & $1.0 \times 10^{9}$ & Kinetic (bipolar) & 0.329 & 0.933 &  DES Y3 3$\times$2pt\\
             SIMBA & $(100 ~h^{-1})^3$ & $1.82 \times 10^{7}$ & Kinetic (bipolar) + X-ray & 0.316 & 0.935 &\textit{Planck }2015\\
             FABLE & $(100~h^{-1})^3$ & $9.51 \times 10^{6}$ & Thermal + Kinetic& 0.590 & 0.991 & \textit{Planck }2018\\
             XFABLE & $(100~ h^{-1})^3$ & $9.51 \times 10^{6}$ & Thermal + Kinetic  & 0.650 &  0.947 & \textit{Planck }2018\\
        \hline
    \end{tabular}
\end{table*}

\subsubsection{BAHAMAS \& BAHAMAS $8.0$} \label{subsubsec:bahamas}
The \bahamas\ simulation suite presented the first attempt at calibrating the feedback models in large-scale cosmological hydrodynamical simulations on observational data of both galaxies and clusters \citep{McCarthy2017}.  The $(400~h^{-1}\mathrm{Mpc})^3$ production volumes have $2\times 1024^3$ particles, with dark matter particles of mass $3.85\times10^9$ ${\mathrm M}_{\rm \odot}/h$ and baryon particles of initial mass 
$7.66\times10^8$ ${\mathrm M}_{\rm \odot}/h$ \citep{McCarthy2017}.  The simulations were run with the OWLS \citep{Schaye2010} version of the \textsc{gadget-3} hydrodynamic code \citep{Springel2005} \citep[see ][for more details]{McCarthy2017}.  A flat $\Lambda$CDM WMAP 9-year cosmology cosmology was adopted \citep{Hinshaw2013}.

Stellar feedback is modelled as kinetic winds according to \citet{Dalla2008}.  This involves assigning a velocity kick to neighbouring gas particles, with the feedback efficiency specified largely by the parameters $v_w$ and $\eta_w$.  $\eta_w$ is the mass loading factor, defined as the ratio of the mass outflow rate to the star formation rate.  The velocity of the outflows are given by $v_w$.  The simulation includes element-by-element radiative cooling \citep{wiersma2009a}, pressure-dependent star formation \citep{schayedalla2008}, and stellar mass loss \citep{wiersma2009b}.

Black hole formation proceeds by placing seed black holes of mass $10^6$~$h^{-1}\, {\rm M}_{\rm \odot}$ into halos with mass greater than $2.75\times 10^{11}$~$h^{-1}\,{\rm M}_{\odot}$.  The black hole accretion rate, $\dot{M}_{\mathrm{BH}}$, is Eddington-limited and proportional to the Bondi-Hoyle-Lyttleton rate \citep{Booth2009}.  A boost-factor $\alpha$ is defined relative to the Bondi–Hoyle accretion rate, and is typically used in some simulations to account for inaccuracies in simulation predictions at high gas density.  These result from current simulations inability to have sufficient resolution to self-consistently track the multi-phase ISM gas, which would result in the Bondi–Hoyle accretion rate being under-estimated if the $\alpha$-factor was neglected. \citet{Booth2009} introduces a density dependent $\alpha$, which is a power-law of the local gas density above a pivot point $n^*_{\mathrm{H, BH}}$. In addition to growing through accretion, black holes can also grow via mergers.  AGN feedback is implemented through the heating scheme described in \citet{Booth2009}, where a fraction of the black hole's bolometric luminosity is coupled thermally to the surrounding gas, resulting in $n_{\mathrm{heat}}$ gas particles having their temperature increased by $T_{\mathrm{heat}}$. 

The parameters of the supernova and AGN feedback subgrid models were calibrated to reproduce the $z=0$ galaxy stellar mass function, the hot gas mass fractions of local groups and clusters, and the amplitude of the present day stellar mass-black hole mass relation.  This resulted in the \bahamas\ production run using $v_w[\mathrm{km s^{-1}}]=300$, $\eta_w=2$
$\log_{10}(T_{\mathrm{heat}}[\mathrm{K}])=7.8$, $n_{\mathrm{heat}}=20$ and $n^*_{\mathrm{H, BH}}[\mathrm{cm}^{-3}])=-1$. 

In addition to using the reference \bahamas\ box with $\log_{10}(T_{\mathrm{heat}}[\mathrm{K}])=7.8$, we also measure the kSZ signal in an otherwise identical simulation, \bahamas\ 8.0, with increased AGN feedback strength; $\log_{10}(T_{\mathrm{heat}}[\mathrm{K}])=8.0$ \citep{McCarthy2017}.  

\subsubsection{FLAMINGO, $f_{\mathrm{gas}}-8\sigma$ \& jets $f_{\mathrm{gas}}-4\sigma$}\label{subsubsec:flamingo}
The \flamingo\ suite \citep{Schaye2023,Kugel2023} is built upon the gravity and hydrodynamics code \textsc{SWIFT} \citep{schaller2023}, incorporating advanced subgrid models for star formation 
\citep{schayedalla2008}, stellar evolution \citep{wiersma2009b}, radiative cooling \citep{ploeckinger2020}, and stellar and AGN feedback. The fiducial simulation (i.e., $\mathrm{L}1\_\mathrm{m}9$) has a box size of 1.0 Gpc with the mass-resolution of baryonic particles of $m_{\mathrm{gas}}$ = $1.0\times10^9~ \mathrm{M}_{\odot}$ and dark matter mass of $m_{\mathrm{DM}}$ = $5.65\times10^9~ \mathrm{M}_{\odot}$. It assumes a flat \(\Lambda\)CDM cosmology from \citet{DESY3All2022},

In the fiducial simulation suite, the thermal AGN feedback model with the modified Bondi-Hoyle black hole accretion is employed as in \citet{Booth2009} and \bahamas\ and kinetic stellar feedback using the prescription of \citet{Chaikin2023}. 
The subgrid model parameters in the suite include two parameters for stellar feedback ($f_{\rm SN}$; multiplicative factor of supernova energy feedback, $\Delta v_{\rm SN}$; wind velocity), one parameter for black hole growth ($\beta_{\rm BH}$; boosted factor for the Bondi-Hoyle accretion rate), and one parameter for the thermal feedback model ($\Delta T_{\rm AGN}$; AGN heating temperature). The parameters are calibrated using machine-learning techniques to match the observed galaxy stellar mass function at $z=0$ in the GAMA survey \citep{Driver2022}. Furthermore, gas mass fractions of groups and clusters are matched to the X-ray observations at $z \sim  0.1$ \citep{Kugel2023}, and the HSC-XXL weak lensing data at $z \sim  0.3$ \citep{Akino_2022}. While the fiducial model has parameter values $f_{\rm SN}=0.238, \Delta v_{\rm SN}=562, \Delta T_{\rm AGN}=10^{7.95}, \beta_{\rm BH}=0.514$, the flamingo suite contains a range of models varying the feedback strength and/or cosmology.  An even broader range of the subgrid physics model parameters can be explored in the \antilles\ suite (Sec.~\ref{subsubsec:antilles}).  

We study additional simulation variants with two different baryonic feedback implementations from \cite{Schaye2023}. In \flamingo\ ``$f_{\mathrm{gas}}-8\sigma$'', the subgrid parameter values of the fiducial AGN feedback models are calibrated to reproduce the cluster gas fraction shifted by -8$\sigma$ from the pre-eROSITA X-ray observations (i.e., $f_{\rm SN}=0.145, \Delta v_{\rm SN}=483, \Delta T_{\rm AGN}=10^{8.40}, \beta_{\rm BH}=0.462$), while keeping the stellar mass function consistent with the fiducial box. Furthermore, we study \flamingo\ ``jets $f_{\mathrm{gas}}-4\sigma$'', where the subgrid parameters are determined with the kinetic, jet-like AGN feedback models \citep{Husko2022} producing the observed cluster gas fraction shifted by -4$\sigma$, while retaining the fit to the $z=0$ stellar mass function (i.e., $f_{\rm SN}=0.176, \Delta v_{\rm SN}=527, v_{\rm jets} =1995\ \rm km/s, \beta_{\rm BH}=0.439$). See Section~2.3.6 of \cite{Schaye2023} for further implementation details.

\subsubsection{SIMBA}\label{subsubsec:simba}
The flagship (100 $h^{-1}$ Mpc)$^3$ \simba\ simulation suite \citep{Dave2019} utilizes 2 $\times$ 1,024$^3$ gas elements, with a gas particle mass of $m_{\mathrm{gas}}$ = $1.82\times10^7~ \mathrm{M_{\odot}}$ and dark matter mass of $m_{\mathrm{DM}}$ = $9.6\times10^7~ \mathrm{M_{\odot}}$. The simulations assume a $\Lambda$CDM cosmology based on \citet{Planck2016}.
\simba\ is built upon the \textsc{Gizmo} code and employs the meshless finite mass (MFM) hydrodynamics scheme \citep{Hopkins2015, Hopkins2017} over the smoothed particle hydrodynamics (SPH) code.

Black hole accretion in \simba\ follows a two-mode approach: torque-limited accretion for cold gas and Bondi accretion for hot gas near the black hole. The use of torque-limited accretion allows a match to observed galaxy-black hole scaling relations in high-resolution simulations without the black hole self-regulating its own growth \citep{HopkinsQuataert2011, Angles_Alcazar2017a}. Additionally, \simba’s bipolar kinetic feedback mechanisms produce both fast and slow outflows depending on the black hole accretion rate (which broadly match the observed trends), along with X-ray radiative feedback model following \cite{Choi2012}. In the flagship \simba\ runs, the model for galactic winds of stellar feedback uses the wind velocity and employs a new mass loading factor based on FIRE simulations \citep{FIRE_Hopkins14, FIRE_Hopkins18, FIRE_Muratov15}.  \simba\ was calibrated to reproduce observations of the stellar mass function, quenched galaxy fractions, and the black hole—stellar mass relation.  Unlike many of the other simulations we study, no gas or halo-scale properties were used in the calibration.


\subsubsection{FABLE \& XFABLE}\label{subsubsec:fable}

We use the $(100~h^{-1}\mathrm{Mpc})^3$ volume  \fable\ simulation model \citep{Henden2018,Bigwood2025}.  The \fable\ $(100~h^{-1}\mathrm{Mpc})^3$ box tracks $1280^3$ dark matter particles with mass $m_{\rm DM} = 3.4 \times 10^7 \, h^{-1} \mathrm{M}_{\odot}$, and $\sim1280^3$ gas cells with a mean target gas cell mass of $\bar{m}_{\rm gas} = 6.4 \times 10^6 \, h^{-1} \mathrm{M}_{\odot}$.  A \citet{Planck2018} cosmology was adopted. 

\fable\ uses the massively-parallel moving mesh hydrodynamic code \textsc{arepo} \citep{Springel2010, Pakmor2016}.  In general, many of the subgrid models, including the star formation implementation \citep{Springel2003}, are identical to those used in the Illustris project \citep{Vogelsberger2013, Torrey2014}.  Alike Illustris, supernova feedback is implemented as a stellar winds model, where wind particles are stochastically launched from star-forming gas with velocities determined by the local dark matter velocity dispersion and are temporarily decoupled from the hydrodynamics.  The \fable\ model however differs in that one-third of the wind energy is thermal, rather than being purely kinetic in Illustris \citep{Marinacci2014, Henden2018}.  As in Illustris, AGN feedback in \fable\ is modelled with two modes, separated by the accretion rate of the black hole \citep{Sijacki2015}.  In the radiatively efficient accretion regime the quasar-mode acts, where a fraction of the available feedback energy is coupled thermally and isotropically to the surrounding gas.  This is typically the dominant process at high redshifts \citep{Martin-Alvarez2024}.   When the black hole is instead accreting in the radiatively inefficient regime, the radio-mode injects bubbles in the gas at a distance from the black hole, in order to mimic the radio lobes inflated by `mechanical' feedback injected by AGN jets.  The feedback parameters are calibrated to match the local galaxy stellar mass function and the gas mass fractions in massive groups and clusters.  

The \xfable\ model was introduced in \citet{Bigwood2025} as an empirical AGN feedback mechanism which exhibits stronger matter power spectrum suppression on large
scales than \fable, while remaining in good agreement with the the galaxy stellar mass function, gas fraction measurements in groups and clusters, and key galaxy cluster X-ray and tSZ scaling relations.  In this work we utilise the $(100~h^{-1}\mathrm{Mpc})^3$ \xfable\ volume, which with the exception of the AGN feedback implementation, has identical resolution, initial conditions and sub-grid models to \fable.  With respect to \fable, \xfable\ has several key differences in the radio-mode model.  Firstly, the hot bubbles are injected at a larger distance of $D_{\mathrm{bub}}=100$~kpc$/h$, rather than $D_{\mathrm{bub}}\approx30$~kpc$/h$ in \fable.  Black holes with larger accretion rates with respect to the Eddington rate of $\dot{M}_{\mathrm{BH}}/\dot{M}_{\mathrm{Edd}}<0.1$ are also able to enter the radio mode, in contrast to $\dot{M}_{\mathrm{BH}}/\dot{M}_{\mathrm{Edd}}<0.01$ in \fable, resulting in a larger population of black holes undergoing radio-mode feedback at a given time.  Thirdly, radio-mode feedback can only operate in host halos $M_{\rm 500} \gtrapprox 10^{13}~\mathrm{M_{\odot}}\,$, i.e. those which have a well-developed `hot atmosphere', with the aim of ensuring the bubbles would be well confined and transfer energy effectively to the ICM.  This produces the sharp feature in the gas and baryon fractions seen in Figure~\ref{fig:simprops}.  In a higher-resolution simulation with the same model, relativistic AGN jet propagation and bubble inflation would be properly resolved, so the location and halo-scale deposition of jet energy would be self-consistently determined, naturally producing scatter in the gas mass fraction relation without introducing abrupt features.  Finally, the energy content of the radio bubble with respect to the ICM energy is limited to $E_{\rm bub}/E_{\rm ICM}<20$, to improve agreement with observations of X-ray cavities that indicate jets-injected bubbles are inflated approximately in pressure-equilibrium \citep{Fabian2012}. 

\fable\ and \xfable\ are part of the larger simulation suite presented in \citet{Bigwood2025}, in which we explore a wide range of feedback parameterizations as a function of cosmic time, host halo properties, and the varying spatial location where feedback energy is deposited.  \xfable\ was calibrated to match pre-eROSITA halo gas fractions. However, recent eROSITA measurements indicating lower halo gas mass fractions \citep{Popesso2024, Dev2024, siegel2025b} suggest that several models previously considered inconsistent should be revisited.

\subsubsection{ANTILLES}\label{subsubsec:antilles}
The \antilles\ simulation suite \citep{Salcido2023} consists of 400 $(100~h^{-1}\mathrm{Mpc})^3$ volumes.  Each box adopts the flat $\Lambda$CDM WMAP 9-year cosmology \citep{Hinshaw2013}. 
The simulations each have resolution matching that utilised in \bahamas\ \citep{McCarthy2017}, corresponding to 256$^3$ baryon and dark matter particles with initial gas mass of $m_{\mathrm{gas}}$ = $1.09\times10^9~ \mathrm{M_{\odot}}$ and dark matter mass of $m_{\mathrm{DM}}$ = $5.51\times10^9~ \mathrm{M_{\odot}}$.  

The simulation suite was run with a modified version of the \textsc{gadget-3} smoothed particle hydrodynamics (SPH) code \citep{Springel2005} created for the \eagle\ project \citep{Schaye2015}.  To explore the impact that the hydrodynamic scheme has on results, 200 of the simulations were run using the standard \textsc{gadget} SPH scheme, and 200 simulations, with otherwise identical initial conditions and sub-grid parameters, were run using the updated \textsc{anarchy} code.  The latter implements the Pressure-Entropy SPH formulation of \citet{Hopkins2013}, in addition to the time-step limiter of \citet{Durier2012}, the thermal diffusion implementation of \citet{Price2008}, and a simplified version of the ‘Inviscid SPH’ artificial viscosity switch of \citet{Cullen2010}.  Further details on the differences between the SPH schemes are discussed in \citet{Schaller2015}. 

We refer the reader to \citet{Salcido2023} and \citet{Schaye2015} for a detailed description of the suite.  Following \bahamas, supernova feedback is implemented as a kinetic wind model \citep{Dalla2008}, black holes grow via the \citet{Booth2009} Eddington-limited boosted Bondi-Hoyle-Lyttleton rate, and AGN feedback follows the \citet{Booth2009} stochastic heating model. The aim of the \antilles\ suite was to explore a wide feedback landscape centered on but going beyond `realistic' scenarios constrained by observations.  As a result, the five subgrid parameters associated with supernova and AGN feedback detailed above were systematically varied between boxes in the \antilles\ simulation suite within the ranges; 

\begin{enumerate}

\item The wind velocity in the kinetic wind stellar feedback model, $v_w[\mathrm{km s^{-1}}]\in[50,350]$.  
\item The mass-loading factor in the kinetic wind stellar feedback model, $\eta_w\in[1,10]$.  
\item  The temperature increase of gas particles in an AGN feedback event, $\log_{10}(T_{\mathrm{heat}}[\mathrm{K}])\in[7,8.5]$.
\item  The number of gas particles heated in an AGN feedback event, $n_{\mathrm{heat}}\in[1,30]$. 
\item The gas density threshold above which the black hole accretion rate is boosted according to the \citet{Booth2009} model, $\log_{10}(n^*_{\mathrm{H, BH}}[\mathrm{cm}^{-3}])\in[-3,-1]$.

\end{enumerate}

The subgrid parameter values were assigned to simulations based on a Latin hypercube sampling with multidimensional uniformity \citep{Deutsch2012}.  As previously mentioned, for each of the 200 unique subgrid parameter combinations, there is a simulation run with the \textsc{gadget} scheme and an analogous box run with \textsc{anarchy}, resulting in a simulation suite of 400 boxes in total.

\subsection{DESI Y1 + ACT kSZ observations}\label{sec:actdata}

The Dark Energy Spectroscopic Instrument (DESI) is a Stage-IV spectroscopic galaxy survey, located at the 4-meter Mayall telescope in Kitt Peak, Arizona.  
One of the primary galaxy types that DESI is targeting are Luminous Red Galaxies (LRGs). LRGs are an excellent galaxy sample for use in stacked measurements of the kSZ effect, not only due to their large luminosities facilitating precise redshift measurements, but also their low satellite fraction makes them less prone to mis-centering with respect to the host halos' hot gas distribution.  Another key DESI sample is the Bright Galaxy Sample (BGS), which covers a range of galaxy-types in the dark energy dominated era, with approximately constant number density in the redshift range $0.1 <z < 0.4$.  The DESI LRG and BGS Y1 samples, obtained after one year of survey operations, overlaps partially with the Atacama Cosmology Telescope Data Release 6 (ACT 6) temperature and polarisation CMB maps.  The maps attained night-time data in three frequency bands (f090: 77–112 GHz, f150: 124–172
GHz and f220:182–277 GHz) between 2017 and 2021, using the 6-metre millimeter waveband ACT instrument based in on Cerro Toco in the Atacama Desert, Chile.  

In this work, we compare our simulation predictions of the kSZ effect to the ACT DR6 + DESI Y1 LRG and BGS kSZ measurements, presented in \citet{Ried2025}.  The measurements stack 825,283 spectroscopically confirmed LRGs at $0.4<z<1.1$, in an area overlapping the two surveys of 4,338 deg$^2$.  The kSZ signal is detected at $9.8\sigma$ significance.  The measurements are reported split into four stellar masses bins: $\log_{10}(M_*/\mathrm{M_{\odot}}) \in [10.5,11.2], [11.2,11.4],[11.4,11.6], [11.6,12.5]$ (LRG M1, LRG M2, LRG M3 and LRG M4, containing 244,932, 320,914, 194,037, and 53,997 LRGs, respectively).  The BGS sample contains 95,934 galaxies and is detected with a signal-to-noise ratio of 2.1.  Throughout the work, we also refer to the the photometric DESI Y1 + ACT LRG measurements in the  $\log_{10}(M_*/\mathrm{M_{\odot}}) \in [11.25,11.50], [11.25,11.50], [11.50,12.0]$ stellar mass bins, referred to here as LRG M1 Photo-z, LRG M2 Photo-z, and LRG M3 Photo-z respectively \citep{Hadzhiyska2024}.  

Following \citet{siegel2025b}, we de-prioritize the LRG M3 and LRG M4 subsamples from the simulation comparison in the main text.  As seen in the right panel of Figure~\ref{fig:tksz_massdep}, their kSZ amplitudes display a differing mass trend at $M_{500} \geq 3\times 10^{13} \mathrm{M_{\odot}}$ relative to the LRG M3 Photo-z measurement, such that the measurements give different conclusions.  Given this uncertainty, we are therefore cautious about drawing robust feedback constraints from the high stellar mass bins. A simulation comparison to the spectroscopic LRG M3 and LRG M4, as well as as LRG M3 Photo-z, is shown in Appendix~\ref{app:highmassbins}.

\section{Methodology for measuring the kSZ effect in hydrodynamical simulations}\label{sec:methods}

Measuring the kSZ effect in simulations in a way that enables robust comparison to observations requires careful treatment of several challenges. Chief among these are:
\begin{itemize}
\item Measurement effects, such as the beam correction (Sections~\ref{sec:kszmaps} and \ref{sec:stackmethod}).
\item Cosmic variance and line-of-sight projections (Section~\ref{sec:cosmicvariance}), which can bias the kSZ amplitude in limited simulation volumes.
\item Sample selection and halo mass calibration, to ensure simulated galaxies are matched to the observed population in terms of their mean halo mass (Section~\ref{sec:sample_selection}).
\item Contributions from satellites (Section~\ref{sec:satellites}), which must be accounted for when stacking galaxy samples.
\item Mis-centering of galaxies relative to their halos (Section \ref{sec:miscentring}), which can dilute the measured signal.
\end{itemize}
In this section we lay out a methodology for addressing each of these challenges, enabling a like-for-like comparison of simulated kSZ profiles to the DESI+ACT measurements.

\subsection{Producing $\Delta T_{\mathrm{kSZ}}$-maps}\label{sec:kszmaps}

Since several of the simulations we study do not generate lightcones on the fly, we construct $\Delta T_{\mathrm{kSZ}}$ maps by manually positioning simulation boxes at the redshift of the snapshot.  By increasing the $z$-coordinates of all particles by the comoving distance to the snapshot's redshift computed for the simulation's cosmology, we position the simulation box so that an observer at the origin would view the box as subtending a small fraction of the celestial sphere.  This allows us to produce a projected $\Delta T_{\mathrm{kSZ}}$ map that mimics a portion of a full-sky light-cone. 

Projecting along the $z$ axis, we compute the Doppler $b$-parameter for gas particles as;
\begin{equation}\label{eq:b}
    b= \frac{v_r \sigma_T m_g}{\mu_e m_p \Omega_{\mathrm{pixel}}d_A^2 c},
\end{equation}

where $m_g$ and $v_r$ are the masses and radial velocities of the gas cells respectively, $\sigma_T$ is the Thomson scattering cross-section, $m_p$ is the mass of a proton and $\mu_e$ is the mean molecular weight per free electron, which we set to $\mu_e=1.17$, assuming a fully ionized plasma of hydrogen and helium, with a helium mass fraction of $Y\approx0.3$.  We use the $z$-component of gas cells' velocities as the radial velocities.  Since the kSZ effect originates from ionized gas, we exclude star forming gas cells and those with temperature $T<10^{5}~\mathrm{K}$.  We verified that using all gas produced negligible changes in the kSZ profile, and therefore are results are not sensitive to this choice.  For the $100~\mathrm{Mpc}/h$ simulation boxes (\antilles, \fable, \xfable\ and \simba), we analogously produce kSZ maps projecting along the $x$ and $y$ axis.  Combining projections along all three independent axes in the smaller simulation volumes reduces the noise that results from small number statistics of halos.  In the case of the larger \bahamas\ and \flamingo\ boxes, using only the $z$ axis is sufficient, as the large number of halos ensures that the noise from halo sampling is negligible.

We approximate $d_A$, the angular diameter distance from the gas cell to the observer, as the updated gas $z$-coordinate at a snapshot redshift minus the box $z$-coordinate of the host halo.  Since we began by increasing the $z$-coordinates of all particles by the comoving distance to the snapshot's redshift, yet all halos are at the same redshift in the box, subtracting the halo position in the box prevents halos with larger $z$ coordinates from increasing $d_A$ and artificially diminishing the $b$-parameter values. 

Assuming a flat-sky approximation at the comoving distance of the simulation box, we linearly transform the $x,y$ coordinates to Right Ascension ($\alpha$) and Declination ($\delta$) values respectively.  We create a square grid in $\alpha$, $\delta$ with cells of equal pixel area, $\Omega_{\mathrm{pixel}}$.  Since the simulations we study have different volumes, which result in the boxes spanning varying fractions of the celestial sphere, we amend the grid created for a given simulation so that an approximately constant resolution of $\Omega_{\mathrm{pixel}}\approx (0.2~ \mathrm{arcmin})^2$ is attained.  We sum the $b$-parameter values of all particles in a given grid cell to produce a projected $b$-map.
 
We transform the $b$-maps to $\Delta T_{\mathrm{kSZ}}$-maps as;
\begin{equation}
    \Delta T_{\mathrm{kSZ}}=-b\Delta T_{\mathrm{cmb}}
\end{equation}
where $T_{\mathrm{cmb}}=2.726$~K is the mean CMB temperature.  

To facilitate comparison with the DESI Y1 + ACT measurements, we convolve the $\Delta T_{\mathrm{kSZ}}$-map with a Gaussian beam with a full width at half maximum (FWHM) equivalent to 1.6 arcmin.

\subsection{Computing the stacked kSZ profile}\label{sec:stackmethod}
Since the kSZ effect is subdominant in CMB maps with respect to the CMB anisotropies, observational detection requires stacking the kSZ signal over a number of galaxies (see Section~\ref{subsubsec:sample_selection} for details on the sample selection).  In the simulations, we thus follow the observational approach of measuring the stacked kSZ profile outlined in \citet{Schaan2021} and \citet{Ried2025}.  This involves applying a compensated aperture photometry (CAP) filter, centered on each galaxy in the sample (see Sec.~\ref{subsubsec:sample_selection} for sample selection), as;
 
\begin{equation}
\mathcal{T}(\theta_\textrm{d})=
\int d^2\theta \, \Delta T_{\rm kSZ}(\theta) \, W_{\theta_\textrm{d}}(\theta) \ \ \ ,
\end{equation}
where the CAP filter is defined as:
\begin{equation}
W_{\theta_\textrm{d}}(\theta) =
\left\{
\begin{aligned}
1& &  &\text{for} \, \theta < \theta_\textrm{d} \,, \\
-1& &  &\text{for} \, \theta_\textrm{d} \leq \theta \leq \sqrt{2}\theta_\textrm{d} \,, \\
0& & &\text{otherwise}. \\
\end{aligned}
\right.
\end{equation}

Within $\theta < \theta_\textrm{d}$, applying the CAP filter has the effect of measuring a cumulative kSZ profile around each galaxy, whilst the subtraction of signal at $\theta_\textrm{d} \leq \theta \leq \sqrt{2}\theta_\textrm{d}$ reduces the impact of noise from CMB fluctuations with longer wavelengths than the filter size.

As the kSZ signal is sensitive to the peculiar velocities of systems, which can be either positive or negative, a simple additive stacking of the kSZ profiles would lead to a cancellation of the signal.  As a result, \citet{Schaan2021} derived a velocity weighted, inverse-variance weighted mean stack of the profiles as the minimum variance unbiased linear estimator:
\begin{equation}
    \hat{T}_{\rm kSZ}(\theta_\textrm{d}) = -
    \frac{1}{r_v}
    \frac{ v_{\rm rms}^{\rm rec}}{  c}
    \frac{\sum_i \mathcal{T}_i(\theta_\textrm{d}) (v_{{\rm rec}, i}/c) / \sigma_i^2}{\sum_i (v_{{\rm rec}, i}/c)^2 / \sigma_i^2}
    \label{eq:kSZ_est}
\end{equation}

The estimator stacks over galaxies, $i$, with peculiar radial velocity $v_{{\rm rec}, i}$ and noise variance for the CAP filter $\sigma_i$, which accounts for noise contributions from the detector and the atmosphere noise, as well as from the primary CMB and foreground effects.  Here $v_{\rm rms}^{\rm rec}$ is the root mean square (RMS) of the peculiar velocities in the galaxy sample and ${r_v}$ is the kSZ bias factor.  This factor accounts for the fact that in \citet{Schaan2021} and \citet{Ried2025}, the peculiar radial velocities are attained through a reconstruction which solves the linearised continuity equation using the galaxy overdensity field.  Since the reconstruction does not account for non-linear effects, in addition to shot noise, there is likely to be a bias between the reconstructed and true radial velocities.  A value of ${r_v}=0.65$ is estimated for the DESI Y1 LRG + ACT sample \citep{Ried2025}.  In the simulations, however, we use the true and therefore unbiased radial velocities, and therefore set  $r_v=1$.  As with the radial velocities of gas particles, we take the $z$-component of galaxies' velocities as the radial velocities.  For satellite galaxies, we adopt the velocity of the host halo's central galaxy. 

We also note that for all stacks, we exclude objects that lie within $6\sqrt{2}$ arcmin of the box edge, to ensure that the CAP filter can capture all gas associated with a given galaxy/group and the kSZ signal is not falsely truncated by the filter hitting the box edge. 


\subsection{Challenge: Mitigating cosmic variance}\label{sec:cosmicvariance}

\begin{figure}
\centering
\includegraphics[width=0.5\textwidth]{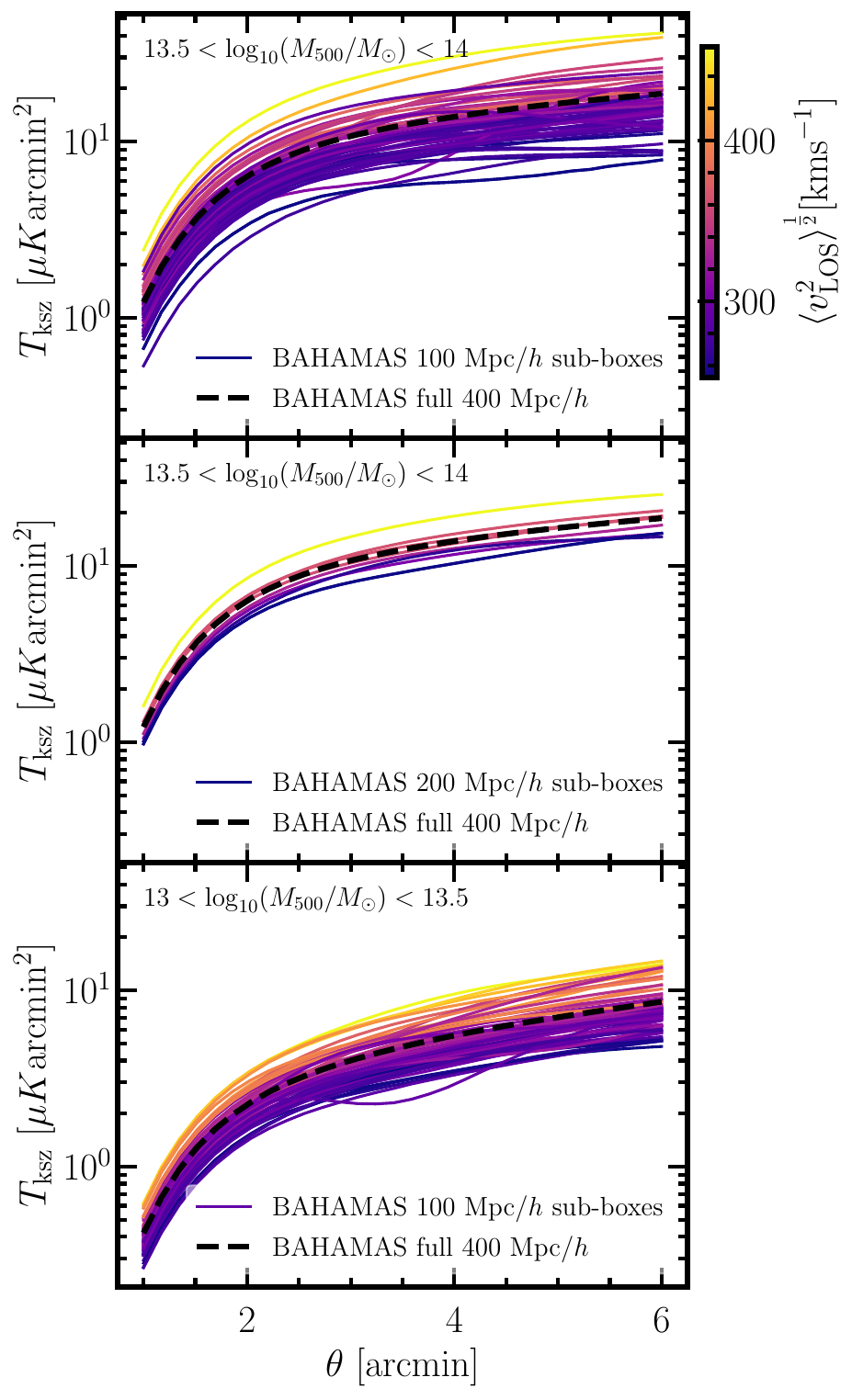}

    \caption{The impact of cosmic variance on the stacked kSZ profile, measured in sub-volumes of the full $400~\mathrm{Mpc}/h$ BAHAMAS box. In all panels, the colour of each sub-volume profile indicates the RMS line-of-sight velocity of halos in the stack, ${\langle v_{\mathrm{LOS}}^2 \rangle}^{1/2}$. The black dashed line shows the result from the full $400~\mathrm{Mpc}/h$ volume.
\textit{Upper}: halos with mass $13.5<\log_{10}(M_{500}/\mathrm{M_{\odot}})<14$ measured in 64 $100~\mathrm{Mpc}/h$ sub-volumes.
\textit{Centre}: halos with mass $13.5<\log_{10}(M_{500}/\mathrm{M_{\odot}})<14$ measured in eight $200~\mathrm{Mpc}/h$ sub-volumes.
\textit{Lower}: halos with mass $13<\log_{10}(M_{500}/\mathrm{M_{\odot}})<13.5$ measured in 64 $100~\mathrm{Mpc}/h$ sub-volumes.}
    \label{fig:cosmicvariance}
\end{figure}

Before attributing differences in the kSZ signal between simulation boxes to feedback, it is important we first understand the scatter that can arise due to cosmic variance alone, especially given the differing volumes of the simulation boxes we study.  To explore this, we measure the stacked kSZ profile in sub-volumes of the full $400~\mathrm{Mpc}/h$ \bahamas\ box.  Figure~\ref{fig:cosmicvariance} shows the stacked kSZ profile for halos with mass $13.5<\log_{10}(M_{500}/\mathrm{M_{\odot}})<14$ measured in the 64 $100~\mathrm{Mpc}/h$ (upper panel) and eight $200~\mathrm{Mpc}/h$ (middle panel) \bahamas\ sub-volumes.  We find that there is significant scatter in the amplitude of the signal measured between different $100~\mathrm{Mpc}/h$ sub-volumes, centred around the profile computed in the full $400~\mathrm{Mpc}/h$ volume.  The size of the scatter decreases by a factor of $\sim2$ when comparing the profile computed in larger $200~\mathrm{Mpc}/h$ sub-volumes.  Similarly, when considering the stacked kSZ profile for halos of mass $13<\log_{10}(M_{500}/\mathrm{M_{\odot}})<13.5$ measured in 64 $100~\mathrm{Mpc}/h$ sub-volumes (lower panel), the scatter in the kSZ amplitude between sub-volumes decreases with respect to the larger halo mass bin. Figure~\ref{fig:cosmicvariance} thus informs us that the scatter arises from limited number statistics (possibly also coupled with the selected galaxies sampling different environments), since larger volumes and lower halo masses increase the number of objects that the kSZ signal is stacked over, and therefore reduces the variance.  Consequently, when analysing the $100~\mathrm{Mpc}/h$ simulation boxes in particular, considerations for cosmic variance will be required before differences in the kSZ amplitude between simulations is attributed to feedback modelling.  Comparing $400~\mathrm{Mpc}/h$ sub-volumes of the \flamingo\ box, or three independent BAHAMAS volumes ran with different random seeds for the initial density field, we see no scatter between the measured kSZ profiles, indicating the box-size effects are converged at that volume.  A similar result, demonstrating that the matter power spectrum suppression is converged in such volumes, is shown in \citet{Schaller2024}.

In searching for the origin of the variance in the kSZ signal between sub-volumes, we found that there was a weak correlation between the mean halo mass of objects stacked in the sub-volume and the amplitude of the kSZ profile, and no trend with tracers of the black hole assembly history.  We found that the key property of sub-boxes that drives the scatter is the RMS of the peculiar velocities of halos stacked to form the kSZ profile, as demonstrated in Figure~\ref{fig:cosmicvariance}.  Limited sample sizes of objects in the sub-box stacks result in the RMS of halo peculiar velocities deviating from the RMS of halo peculiar velocities measured for the full simulation volume. Since the magnitude of the kSZ signal depends linearly on the radial velocities of halos (Equation~\ref{eq:b}), the scatter in the RMS line of sight velocity of halos included in the kSZ stack in different sub-volumes results in the variations in the measured amplitude of the kSZ signal. 

Given that this effect arises due to limited sample statistics and results in a significant scatter in the kSZ amplitude in $100~\mathrm{Mpc}/h$ volumes, we hereafter apply a correction to the stacked kSZ profiles measured in each simulation to account for the effect of cosmic variance.   For each measurement of the kSZ profile in a given simulation and halo mass/stellar mass selection, we compute the RMS value of the radial velocity of galaxies in the stack, ${\langle v_{\mathrm{LOS}}^2 \rangle}^{\frac{1}{2}}$.  We then correct the amplitude of each profile by the ratio of the RMS value of the radial velocity of galaxies in the stack measured in FLAMINGO to the value measured in the simulation of interest.  With this correction factor, we i) account for scatter in ${\langle v_{\mathrm{LOS}}^2 \rangle}^{\frac{1}{2}}$ between boxes due to cosmic variance effects in small volumes and ii) effectively rescale all simulations to the FLAMINGO cosmology, allowing differences in the kSZ profiles between simulations to be interpreted as arising solely from differences in the modelling of galaxy formation.  For our key comparison of simulated kSZ profiles to the DESI Y1 + ACT measurements, we show the equivalent of Figures~\ref{fig:tksz_ggl_main} and \ref{fig:tksz_massdep} with uncorrected ${\langle v_{\mathrm{LOS}}^2 \rangle}^{\frac{1}{2}}$ in Appendix~\ref{fig:unscaled}.

Another consequence of cosmic variance that we observe is the dependence of the kSZ signal’s amplitude and shape on the choice of line-of-sight axis, which is again driven by small number statistics.  As discussed in Section~\ref{sec:kszmaps}, to mitigate this in the $100~\mathrm{Mpc}/h$ simulation boxes (\antilles, \fable, \xfable, and \simba), we compute the kSZ signal by stacking over the three independent projections along the $x$, $y$, and $z$ axes.  When comparing to the DESI + ACT data, we further show a shaded region for \fable, \xfable\ and \simba, representing the span in the measured kSZ profile attained by projecting along each of the three independent box axes. For the larger simulation boxes, the differences between axis projections are negligible, and therefore we do not show them.

\subsection{Challenge: Galaxy selection}\label{sec:sample_selection}

As emphasised in \citet{McCarthy2024} and \citet{siegel2025b}, in order to enable a like-for-like comparison of kSZ measurements in simulations and observations, we would ideally ensure that the halo mass distributions of samples are well matched.  However, since halo masses are not directly observable, the DESI Y1 + ACT measurements are reported in bins of stellar mass \citep{Ried2025}.  Since the stellar mass-halo mass relation varies between different hydrodynamical simulations and is not necessarily calibrated to observations, and the observations themselves are subject to uncertainty, there is no guarantee that a realistic mapping to halo mass would occur if one simply selected stellar mass bins in the simulations.  Furthermore, a simple stellar mass cut would not ensure that a realistic satellite fraction is maintained in the simulations.  
In the absence of accurate halo mass estimates, a standard approach is to apply the abundance matching formalism which links an observable quantity (such as stellar mass) with halo mass at fixed abundance.  In the following section, we show that this approach can lead to non-negligible biases in the inferred halo mass due to differences in the stellar mass-halo mass relation between simulations.

Following \citet{McCarthy2024}, we do not attempt to reproduce a full DESI LRG selection of galaxies in the simulations; a non-trivial task given the complexity and uncertainties involved in mapping intrinsic stellar properties to observed luminosities. However, we expect that any kSZ stack is primarily sensitive to the mean halo mass, owing to the approximately linear relations between kSZ signal, gas mass, and halo mass. We refer the reader to Appendix A of \citet{McCarthy2024}, which demonstrates that applying a more stringent selection which imposes that galaxies are non-star-forming has a negligible impact on the kSZ profile.


\subsubsection{Selection via abundance matching}\label{subsubsec:sample_selection}

In the simulations, we rank order galaxies in descending order by stellar mass, and compute the cumulative number density.  We define a galaxy's stellar mass by summing stellar particles within a spherical aperture of radius  50 pkpc centered on the subhalo particle with the minimum gravitational potential.  Using the observational GSMF measurements presented in \citet{Behroozi2019} at the closest redshift to the simulation snapshot, we integrate the GSMF to similarly obtain the cumulative number density of galaxies above a given stellar mass.  Via an abundance matching approach, we then reassign galaxies in the simulation stellar masses based on matching the cumulative number densities between the observations and simulations.  Using the re-assigned stellar masses, we select galaxies in the simulations based on the reported DESI Y1 + ACT stellar mass bins: $\log_{10}(M_{*}/\mathrm{M_{\odot}}) \in \mathrm{LRG\ M1:}[10.5, 11.2], \mathrm{LRG\ M2:}[11.2, 11.4],\mathrm{LRG\ M3:}[11.4, 11.6],$ $\mathrm{LRG\ M4:}[11.6, 12.5]$, where the stellar mass estimates for the DESI LRG sample were obtained from DESI legacy survey photometry in \citet{zhou2023}. 

Figure~\ref{fig:m500_mstar_selected} shows the mean halo mass, $ \log_{10}(\left\langle M_{500}[\mathrm{M_{\odot}}]\right\rangle)$, of galaxies selected in each stellar mass bin via the abundance matching approach, for each mass bin for each simulation.  We find that the mean halo mass measured for a given stellar mass bin varies significantly between simulations, resulting from differences in the stellar mass-halo mass relation.  Hence, since an abundance matching method based on stellar masses in the simulations does not guarantee that it will provide samples of equivalent halo mass that can be compared on a like-for-like basis, it is challenging to use this method when comparing the kSZ signal from simulations with observations. In the following sections, therefore, we adopt the approach used by \citet{McCarthy2024} and \citet{siegel2025b} to select galaxies from simulations.

\subsubsection{Selection via galaxy-galaxy lensing}\label{subsubsec:sample_selectionggl}

As shown by \citet{McCarthy2024} and \citet{siegel2025b}, one way to select simulation samples with halo masses that are well-matched to observations is to measure the total matter profiles of the galaxies used in kSZ stacks via weak gravitational lensing.  \citet{siegel2025b} measured the GGL signal around DESI Y1 BGS and LRG objects in each of the stellar mass bins reported in \citet{Ried2025}.  By varying the minimum stellar mass, galaxy samples were tuned in the \flamingo\ simulations to fit the projected matter density obtained through galaxy-galaxy lensing in the DESI samples. This procedure ensures that the sample selections between data and simulations are matched in the mean halo mass (see Figure 2 in \citet{siegel2025b}).  In this paper, for each hydrodynamical simulation, we determine the minimum stellar mass cut required to match the mean halo mass in each DESI + ACT kSZ effect bin to the best-fit value from the fiducial \flamingo\ simulation reported in \citet{siegel2025b}. We then stack all galaxies above this minimum stellar mass threshold and measure the resulting kSZ profile.  The required minimum stellar mass cut for each simulation to match the BGS and LRG full samples, as well as the stellar mass binned samples, are summarized in Tables~\ref{tab:sim_ksz_stats_ggl} and \ref{tab:sim_ksz_stats_ggl_highmass}.

\subsection{Challenge: satellite contributions}\label{sec:satellites}

\begin{figure}
\centering

\includegraphics[width=0.4\textwidth]{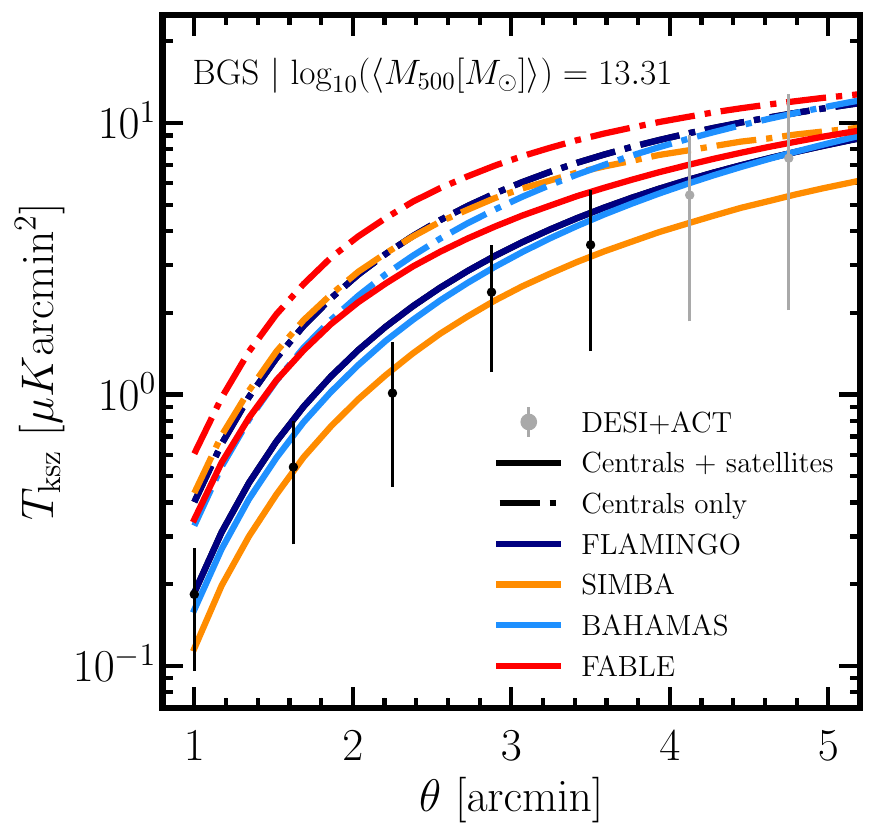}
    \caption{The impact of a centrals-only galaxy selection (dashdotted lines), versus a selection including all galaxies and therefore a fraction of satellite galaxies (solid lines), for BGS-like samples at $z \approx 0.3$ with fixed mean halo mass of $\log_{10}\left(\langle M_{500}[\mathrm{M_{\odot}}] \rangle\right) = 13.30$.  We show 
    \flamingo\ (navy), \simba\ (orange), \bahamas\ (blue) and \fable\ (red). }
    \label{fig:satellite}
\end{figure}

A realistic mock DESI BGS or LRG galaxy sample should include a fraction of satellite galaxies.  However, analytic models for interpreting the kSZ signal typically assume that all galaxies lie at the centre of their host halo. As discussed in \citet{McCarthy2024}, neglecting satellites introduces two competing effects. Firstly, satellites typically reside in more massive halos than centrals of the same stellar mass, which can increase their kSZ signal.  However, because we match the mean halo masses of the simulated and observed samples, our results are not affected by this bias.  Secondly, given satellites are offset from the centers of their host halos' hot gas distribution, their kSZ signal can be reduced with respect to a central galaxy. This is demonstrated in Figure~\ref{fig:satellite}, which compares centrals-only galaxy selections (dash-dotted lines) to one that includes satellites (solid lines), for BGS-like samples at $z \approx 0.3$ with a fixed mean halo mass of $\log_{10}\left(\langle M_{500}[\mathrm{M_{\odot}}] \rangle\right) = 13.31$.  For \flamingo, \simba, \bahamas\ and \fable, a centrals-only selection yields a larger kSZ amplitude.   

In our stellar mass–based selection from hydrodynamical simulations, we naturally include satellites in our mock samples. 
Depending on the mass bin and simulation, we find satellite fractions of $\sim$10-30\% (see Table~\ref{tab:sim_ksz_stats_ggl}). While there remains uncertainty in the satellite fractions of the observed DESI samples, our results are broadly consistent with the values reported in literature \citep{Yuan2024}.

\subsection{Challenge: mis-centering}\label{sec:miscentring}

Observed galaxies are not necessarily perfectly centered on the hot gas distribution of their host haloes. Analytic models for interpreting the kSZ signal, however, commonly assume that galaxies are exactly centered on the halo, which can bias the inferred extent of the gas distribution. In contrast, self-consistent cosmological hydrodynamical simulations naturally produce galaxies that are mis-centred relative to the hot gas, as seen in observations. Mis-centering is therefore not a major source of concern in our analysis.

Another subdominant source of mis-centering that simulations are however susceptible to arises from the uncertainty in galaxy positions due to the $0.5$~arcmin pixel size of the ACT maps. To test the impact of this in the simulations, we displaced the CAP filter centre from the galaxy position, drawing random offsets in both $\alpha$ and $\delta$ from a uniform distribution with a width of 0.5 arcmin. We found that this had a negligible impact on the shape of the inner kSZ profile. 

\section{Results: simulated profiles of the kSZ effect}\label{sec:measurement}
In this section, we first  build intuition on the relationship between the baryon feedback model and the stacked kSZ profiles from predictions using a centrals-only galaxy selection in four bins of halo mass (Section~\ref{subsec:halomass_dep}). Second, we present a like-for-like comparison of the simulations to the DESI Y1 + ACT BGS and LRG  measurements, using a galaxy-galaxy lensing informed sample selection (Section~\ref{subsec:comparison}) and we test the mass dependence by studying the LRGs binned by stellar mass (Section~\ref{subsec:stellarmass_dep}). Finally, we analyze the \antilles\ hypercube suite to find the simulation that best describes the kSZ measurements (Section~\ref{sec:antillescompare}). 


\subsection{The kSZ effect across simulations: centrals-only}\label{subsec:halomass_dep}

\begin{figure*}
\centering
\includegraphics[width=\textwidth]{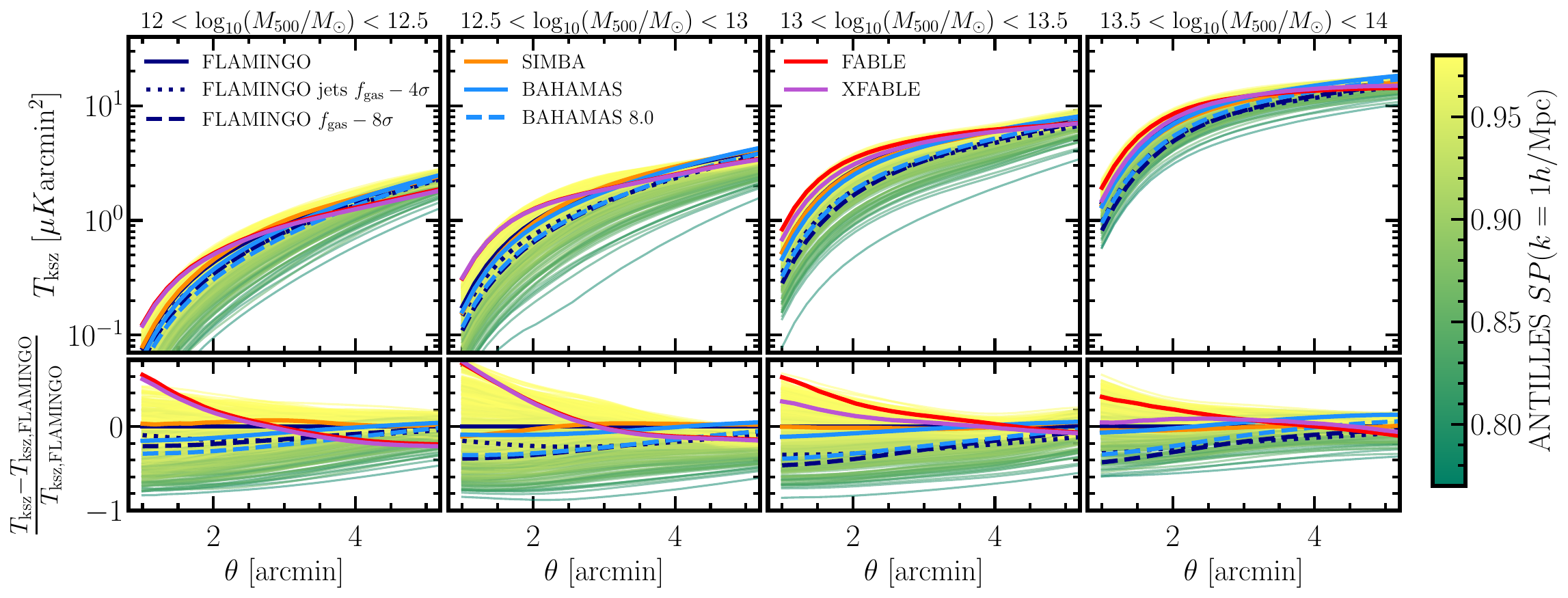}\\
    \caption{The halo mass dependence of the stacked kSZ radial profile in hydrodynamical simulations at $z=0.75$.  We measure the kSZ signal within a CAP filter centered on groups and clusters, stacking the profiles in four $M_{500}$ bins. We plot the 400 \antilles\ boxes, as well as \fable\ (red), \xfable\ (purple), \flamingo\ (navy solid), \flamingo$f_{\mathrm{gas}}-8\sigma$ (navy dashed), \flamingo\ jets $f_{\mathrm{gas}}-4\sigma$ (navy dotted), \bahamas\ (blue solid), \bahamas\ 8.0 (blue dashed) and \simba\ (orange).  The \antilles\ simulations are colour-coded by the suppression of the matter power spectrum due to baryonic effects measured at $z=0,\ k=1~\mathrm{Mpc}/h$. To better distinguish between simulations, the residual panel shows the difference in the kSZ profile of each simulation relative to \flamingo.}
    \label{fig:halo_mass_dep}
\end{figure*}


Figure~\ref{fig:halo_mass_dep} shows the stacked kSZ radial profiles of halos in four $M_{500}$ bins: $\log_{10}(M_{500}/\mathrm{M_{\odot}}) \in [12,12.5], [12.5,13],[13,13.5], [13.5,14]$, for central galaxies at $z=0.75$, predicted from the simulations. Table~\ref{tab:sim_ksz_stats_halo} reports the mean halo mass in each bin, the RMS line-of-sight velocity used to correct for cosmic variance (Section~\ref{sec:cosmicvariance}) and the number of objects. In all simulations, the amplitude of the kSZ signal increases with halo mass, consistent with the expectation from equation~\ref{eq:b}, where $\Delta T_{\mathrm{kSZ}}$ scales with the total halo gas mass, and therefore with halo mass. We echo the finding of \citet{siegel2025b}, that differences in both the amplitude and the shape of the kSZ profile exhibits a mass-dependence.
Because the mean halo masses are well matched across the simulations, and cosmic variance is accounted for, differences in the predicted kSZ amplitudes can be attributed to feedback physics (aside from the small differences in $\Omega_b/\Omega_m$ between simulations).

By comparing the simulation variants within the \bahamas\ and \flamingo\ suites, we can study the sensitivity of kSZ to baryonic feedback strength.
The stronger feedback variants, \bahamas\ 8.0, \flamingo$f_{\mathrm{gas}}-8\sigma$ and \flamingo\ jets $f_{\mathrm{gas}}-4\sigma$ (i.e. those that exhibit an increased suppression of the matter power spectrum) produce lower kSZ amplitudes than their fiducial model across all halo-mass bins, reflecting reduced gas and baryon fractions (Figure~\ref{fig:simprops}). 
We find that the feedback variants converge to approximately the same values for the $T_{\mathrm{kSZ}}$ signal at $\theta=6$\ arcmin, indicating that the halo gas is not removed, but is redistributed to larger distances in the strong feedback simulations. 

The \antilles\ suite show a consistent pattern. Since all 400 boxes share initial conditions and differ only in feedback parameter choices, the scatter directly reflects how subgrid prescriptions redistribute gas to large radii. The resulting variation in profile amplitude provides a clean demonstration that the kSZ responds to feedback efficiency: the amplitude of the kSZ signal in a given halo mass bin correlates with the baryonic suppression of the matter power spectrum measured at $z=0,\ k=1~\mathrm{Mpc}/h$. We study this further in Section~\ref{sec:kszsummary}.

By comparing the \fable\ and \xfable\ predictions, we gain insight into the nuances of feedback modeling. These simulations differ only in their subgrid AGN feedback model (see Section~\ref{subsubsec:fable}). Their profiles are nearly identical in the two lowest mass bins, but diverge at higher masses: \xfable\ gives a shallower inner profile at $\theta<3~\mathrm{arcmin}$ with respect to \fable.  Given that \xfable\ exhibits a greater suppression of the matter power spectrum than \fable\ at $z\leq2$, this deviates from the trend observed for the \antilles, \bahamas\ and \flamingo\ suites, that a larger matter power spectrum suppression correlates with a lower kSZ amplitude across all masses. This reflects the long-range kinetic mode in \xfable, which injects bubbles farther from the black hole once halos exceed $M_{\rm 500} \gtrapprox 10^{13}~\mathrm{M_{\odot}}$. This mode injects more gas in the form of radio bubbles at a greater $r_{500}$, reducing gas fractions and lowering the kSZ amplitude only at high mass. Therefore, despite \xfable\ showing stronger matter power suppression than \fable\ (and assuming the suppression is converged at this volume), the expected monotonic relation between suppression and kSZ amplitude breaks down; a reminder that the mapping depends on how feedback is implemented, not only on its overall strength.  

Comparing all simulations, the \antilles\ kSZ profiles bracket the range of predictions. The inner kSZ amplitude ($\theta<2$~arcmin), which traces the gas within $r_{500}$, matches the relative ordering of gas mass fractions (measured within $r_{500}$) in Figure~\ref{fig:simprops}, providing a key consistency check.
Moreover, the qualitative trends in radial gas profiles (Figure~\ref{fig:simprops}) are reproduced in the kSZ, confirming that both probe the extended gas. 
The strong feedback variants, \flamingo$f_{\mathrm{gas}}-8\sigma$, \flamingo\ jets $f_{\mathrm{gas}}-4\sigma$, and \bahamas\ 8.0, which have lower gas mass fractions than the fiducial X-ray benchmark observations, exhibit the lowest-amplitude kSZ profiles across all four halo mass bins, consistent with their depleted gas fractions.  
Both \fable\ and \xfable\ display flatter profiles, suggesting less extended hot gas distributions, in agreement with their radial gas profiles.

Taken together, the kSZ amplitude increases more rapidly with radius at higher halo masses, and its normalization and shape are strongly determined by feedback physics. Stronger feedback lowers the signal, primarily by redistributing gas to larger radii. The  agreement with gas fraction measurements underscores the robustness of the kSZ as a diagnostic of baryonic feedback, while its sensitivity to feedback implementation, such as kinetic versus thermal AGN modes, demonstrates that the kSZ captures additional information beyond what halo gas fractions alone can provide.

\subsection{Comparison to the DESI + ACT Y1 observations: preference for strong feedback models}\label{subsec:comparison}

\begin{figure}
\centering
\includegraphics[width=\columnwidth]{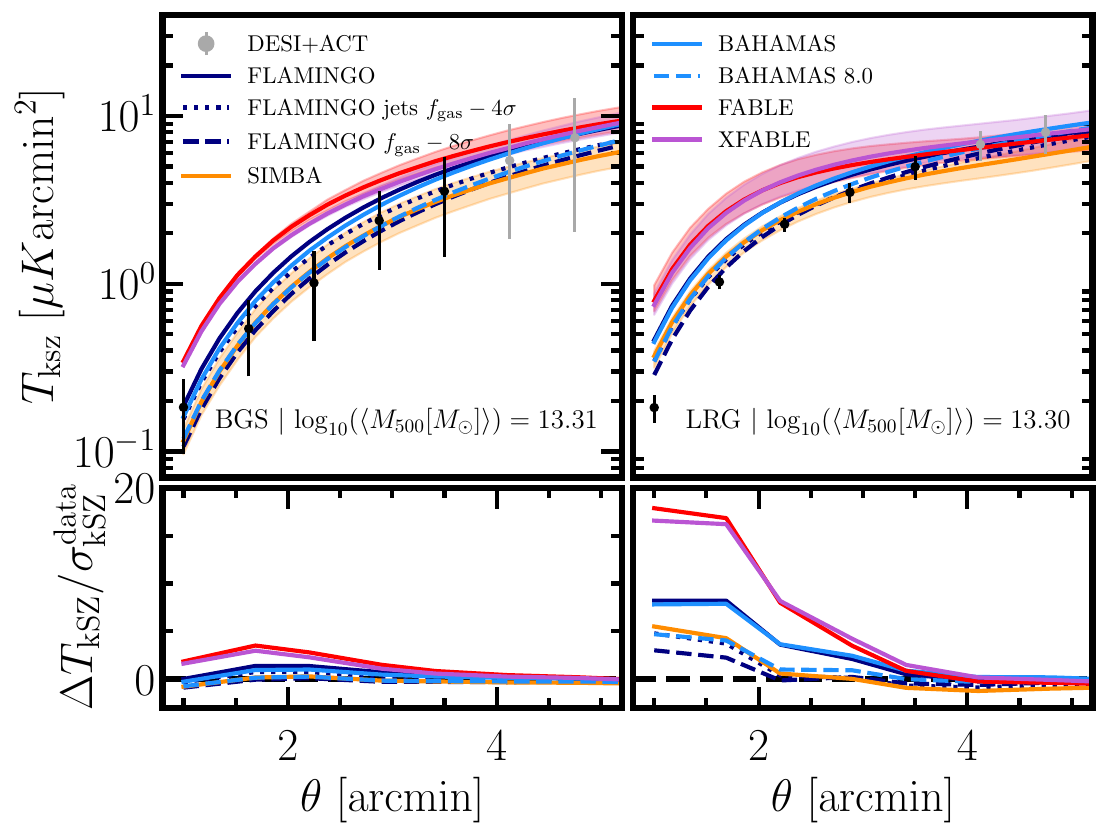}

    \caption{\textit{Upper panels:} the stacked radial kSZ profile in hydrodynamical simulations in comparison to the DESI Y1 + ACT  $z\approx0.3$ BGS and $z\approx0.75$ full sample LRG measurements \citep[black data points;][]{Ried2025}.  We shade the measurements at large radii in grey, since they are correlated and therefore effectively do not provide independent constraints.  We measure the kSZ signal in simulations within a CAP filter centered on galaxies, with galaxy samples constructed using a stellar mass–based selection, ensuring that the mean halo masses, $\log_{10}(\langle M_{\rm 500}[\mathrm{M_{\odot}}]\rangle)$, match those estimated from the galaxy–galaxy lensing measurements reported in \citet{siegel2025b}. We plot \fable\ (red), \xfable\ (purple), \flamingo\ (navy solid), \flamingo$f_{\mathrm{gas}}-8\sigma$ (navy dashed), \flamingo\ jets $f_{\mathrm{gas}}-4\sigma$ (navy dotted), \bahamas\ (blue solid), \bahamas\ 8.0 (blue dashed) and \simba\ (orange).  We show a shaded region for the 100~Mpc/$h$ boxes, representing the span in the measured kSZ profile attained by projecting along each of the three independent box axes, to demonstrate the impact of cosmic variance.  \textit{Lower panels:} the difference between the simulation predicted kSZ signal and the DESI Y1 + ACT measurements, normalised by the reported DESI Y1 + ACT error bars.  }
    \label{fig:tksz_ggl_main}
\end{figure}

\begin{figure*}
\centering
\includegraphics[width=0.9\textwidth]{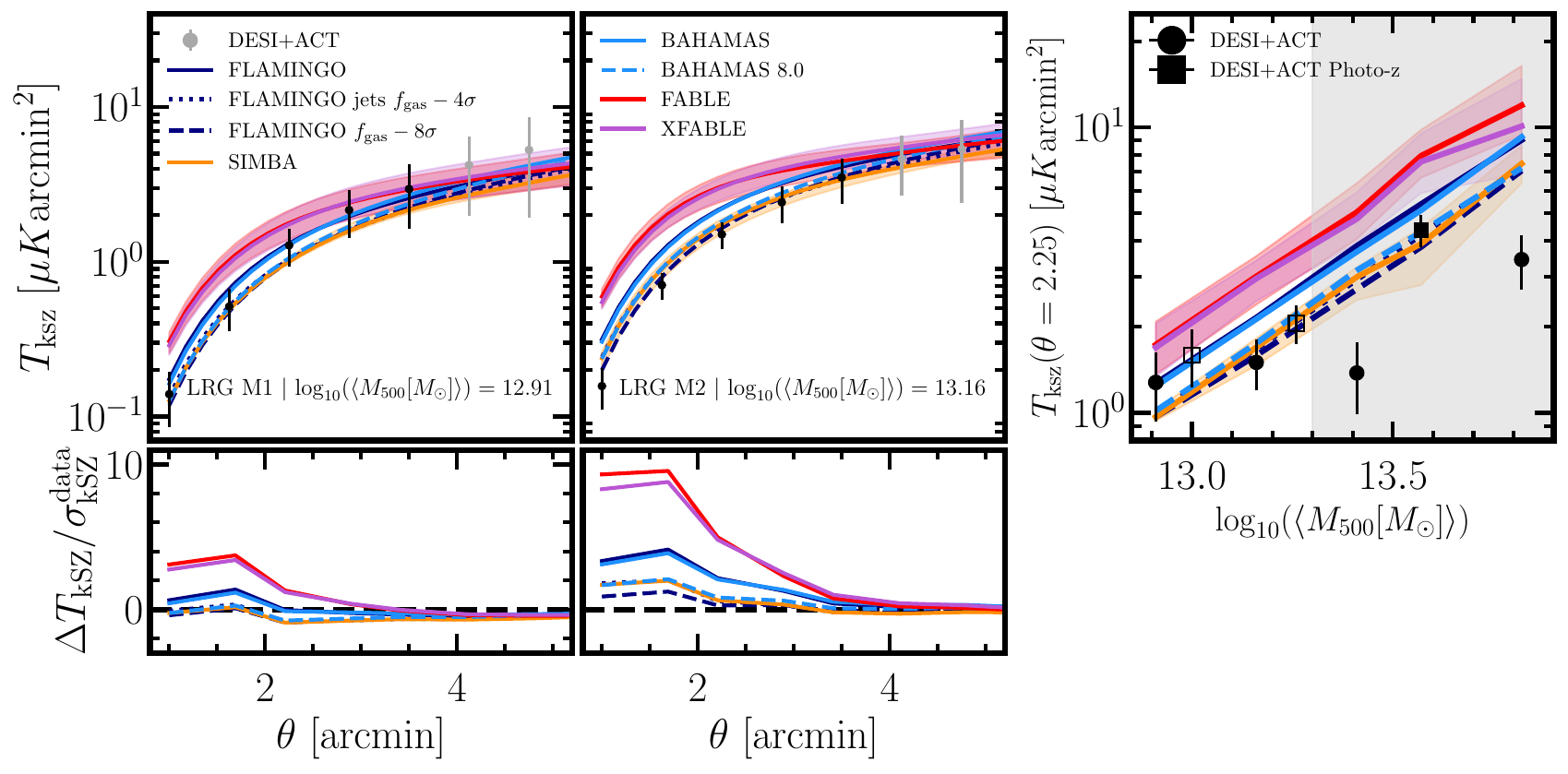}
    \caption{The mass dependence of the kSZ effect. \textit{Left:} the stacked radial kSZ profile in hydrodynamical simulations in comparison to the $z\approx0.75$ DESI + ACT LRG measurements in the M1 ($\log_{10}(M_*/\mathrm{M_{\odot}}) \in [10.5,11.2]$) and M2 ($\log_{10}(M_*/\mathrm{M_{\odot}}) \in [11.2,11.4]$) stellar mass bins \citep[black data points;][]{Ried2025}.  We shade the measurements at large radii in grey, since they are correlated and are therefore effectively not providing independent constraints.  We measure the kSZ signal in simulations within a CAP filter centered on galaxies, with galaxy samples constructed using a stellar mass–based selection, ensuring that the mean halo masses, $ \log_{10}(\langle M_{\rm 500}/\mathrm{M_{\odot}}\rangle)$, in each bin match those estimated from the galaxy–galaxy lensing measurements reported in \citet{siegel2025b}. We plot \fable\ (red), \xfable\ (purple), \flamingo\ (navy solid), \flamingo$f_{\mathrm{gas}}-8\sigma$ (navy dashed), \flamingo\ jets $f_{\mathrm{gas}}-4\sigma$ (navy dotted), \bahamas\ (blue solid), \bahamas\ 8.0 (blue dashed) and \simba\ (orange).  We show a shaded region for the 100~Mpc/$h$ boxes, representing the span in the measured kSZ profile attained by projecting along each of the three independent box axes, to demonstrate the impact of cosmic variance.  The residual panels show the difference between the simulation predicted kSZ signal and the DESI Y1 + ACT measurements, normalised by the reported DESI Y1 + ACT error bars.  \textit{Right:} The mass dependence of the amplitude of the kSZ 
    profile at $z=0.75$, measured at $\theta=2.25~\mathrm{arcmin}$.  We compare to the spectroscopic DESI Y1 + ACT measurements \citep{Ried2025} and the photometric DESI Y1 + ACT LRG M1, M2 and M3 bins \citep{Hadzhiyska2024}.  Following the discussion in Section~\ref{sec:actdata} we shade the region $\log_{10}(M_{500}/\mathrm{M_{\odot}})>13.3$ and do not make simulation measurements for the photometric DESI Y1 + ACT LRG M1 and M2 bins, so show those data points unfilled. }
    \label{fig:tksz_massdep}
\end{figure*}


\begin{table*}
\centering
\caption{Properties of the galaxy samples selected in each simulation for a like-for-like comparison with the stacked kSZ signals from DESI Y1 + ACT \citep{Ried2025}.  We report the minimum stellar mass limit used to select the sample ($\log_{10}(\langle M_{*, \mathrm{min}}\rangle)$), the mean halo mass ($\log_{10}(\langle M_{500}[\mathrm{M_{\odot}}]\rangle)$), the root-mean square of line-of-sight velocities (${\langle v_{\mathrm{LOS}}^2 \rangle}^{\frac{1}{2}}$), the number of galaxies in the stack ($N_{\rm stack}$), the satellite fraction ($f_{\rm sat}$) and the number of standard deviations by which the simulations deviate from the observations ($\sigma$). For all the simulations except the FLAMINGO and BAHAMAS suites, $N_{\rm stack}$ is the sum of the number of halos used in all the three projections along the line-of-sight axis, and the other quantities are the average of the quantities over the three axes. \label{tab:sim_ksz_stats_ggl}}

\begin{tabular}{@{}c c @{} cccccc}
\hline \hline
Bin & Simulation & $\log_{10}(\langle M_{*, \mathrm{min}}[\mathrm{M_{\odot}}]\rangle)$ & $\log_{10}(\langle M_{500}[\mathrm{M_{\odot}}]\rangle)$  & ${\langle v_{\mathrm{LOS}}^2 \rangle}^{\frac{1}{2}}$ [km/s] & $N_{\rm stack}$ & $f_{\rm sat}$ [\%] & $\sigma$\\
\hline
\multirow{9}{*}{BGS} 
  & ANTILLES & 10.67 - 11.16 & 13.31 - 13.33 & 190.4 - 211.6 & 926 - 6922 & 15.9 - 24.9  & 0.60 - 4.56 \\
  & BAHAMAS & 11.10 & 13.31 & 275.59 & 93153 & 21.6 & 1.31 \\
  & BAHAMAS 8.0 & 11.05 & 13.31 & 275.19 & 65583 & 20.5 & 0.91 \\
  & FLAMINGO & 10.94 & 13.31 & 310.26 & 588141 & 26.4 & 1.54 \\
  & FLAMINGO $f_{\rm gas} - 8\sigma$ & 11.01 & 13.31 & 310.51 & 486381 & 25.6 & 0.88 \\
  & FLAMINGO jets $f_{\rm gas} - 4\sigma$ & 10.98 & 13.31 & 309.74 & 382830 & 24.3 & 1.06 \\
  & SIMBA & 10.97 & 13.33 & 219.37 & 10889 & 29.7 & 0.87 \\
  & FABLE & 10.76 & 13.32 & 225.56 & 7086 & 23.5 & 3.28 \\
  & XFABLE & 10.87 & 13.32 & 224.91 & 6232 & 23.7 & 2.75 \\
\cline{1-8}
\multirow{9}{*}{LRG} 
 & ANTILLES & 10.82 - 11.45 & 13.29 - 13.34 & 176.9 - 207.2 & 177 - 1457 & 10.0 - 20.8 & 2.11 - $>10$ \\
  & BAHAMAS & 11.24 & 13.30 & 269.04 & 15839 & 15.1 & $>10$ \\
  & BAHAMAS 8.0 & 11.25 & 13.30 & 268.98 & 11373 & 14.6 & 4.46 \\
  & FLAMINGO & 11.18 & 13.30 & 297.76 & 138262 & 20.2 & $>10$ \\
  & FLAMINGO $f_{\rm gas} - 8\sigma$ & 11.23 & 13.30 & 298.00 & 116947 & 19.7 & 2.90 \\
  & FLAMINGO jets $f_{\rm gas} - 4\sigma$ & 11.14 & 13.30 & 297.38 & 99071 & 19.2 & 4.54 \\
  & SIMBA & 10.98 & 13.31 & 198.06 & 1446 & 16.9 & 5.46 \\
  & FABLE & 11.08 & 13.32 & 229.95 & 1383 & 14.8 & $>10$ \\
  & XFABLE & 11.19 & 13.32 & 222.86 & 1035 & 14.3 & $>10$ \\
\cline{1-8}
\multirow{9}{*}{LRG M1} 
  & ANTILLES & 10.22 - 10.73 & 12.90 - 12.93 & 183.36 - 207.35 & 1862 - 14123 & 11.0 - 23.6 & 0.10 - 2.09 \\
  & BAHAMAS & 10.52 & 12.91 & 264.92 & 290963 & 22.7 & 0.14 \\
  & BAHAMAS 8.0 & 10.55 & 12.91 & 264.97 & 241561 & 22.0 & 0.10 \\
  & FLAMINGO & 10.41 & 12.91 & 296.07 & 1833082 & 28.8 & 0.20 \\
  & FLAMINGO $f_{\rm gas} - 8\sigma$ & 10.55 & 12.91 & 296.09 & 1660547 & 28.4 & 0.11 \\
  & FLAMINGO jets $f_{\rm gas} - 4\sigma$ & 10.58 & 12.91 & 295.87 & 1418923 & 27.5 & 0.11 \\
  & SIMBA & 10.30 & 12.93 & 211.00 & 18232 & 27.7 & 0.10 \\
  & FABLE & 10.22 & 12.94 & 218.04 & 15275 & 23.8 & 2.50 \\
  & XFABLE & 10.29 & 12.94 & 217.40 & 13746 & 23.3 & 2.06 \\
\cline{1-8}
\multirow{9}{*}{LRG M2} 
  & ANTILLES & 10.69 - 11.24 & 13.14 - 13.18 & 179.71 - 206.91 & 344 - 3214 & 11.4 - 21.5 & 0.94 - $>10$ \\
  & BAHAMAS & 11.05 & 13.16 & 267.75 & 44102 & 17.6 & 2.78 \\
  & BAHAMAS 8.0 & 11.07 & 13.16 & 266.91 & 33558 & 16.7 & 1.35 \\
  & FLAMINGO & 11.02 & 13.16 & 297.23 & 321261 & 23.2 & 3.01 \\
  & FLAMINGO $f_{\rm gas} - 8\sigma$ & 11.08 & 13.16 & 297.36 & 274149 & 22.7 & 0.88 \\
  & FLAMINGO jets $f_{\rm gas} - 4\sigma$ & 11.01 & 13.16 & 297.19 & 227874 & 22.0 & 1.34 \\
  & SIMBA & 10.78 & 13.18 & 204.39 & 3932 & 23.1 & 1.20 \\
  & FABLE & 10.87 & 13.18 & 225.04 & 2926 & 17.4 & $>10$ \\
  & XFABLE & 10.97 & 13.18 & 224.77 & 2537 & 17.8 & 8.12 \\

\hline \hline
\end{tabular}

\end{table*}

\begin{table}
\centering
\caption{The combined number of standard deviations by which each simulation prediction deviates from the observations, computed using the ACT + DES BGS and LRG M1 and M2 stellar mass binned measurements, $\sigma_{\rm BGS+M1+M2}$.} \label{tab:sigma}

\begin{tabular}{ccc}
\hline \hline
Simulation &  $\sigma_{\rm BGS+M1+M2}$ \\
\hline
BAHAMAS & 3.1   \\
BAHAMAS 8.0  & 1.6  \\
FLAMINGO &  3.4  \\
FLAMINGO $f_{\rm gas} - 8\sigma$ &  1.2  \\
FLAMINGO jets $f_{\rm gas} - 4\sigma$ & 1.7 \\
SIMBA & 1.5  \\
FABLE &  $>$10 \\
XFABLE &  8.7  \\
\hline \hline
\end{tabular}

\end{table}

In this subsection we compare the stacked kSZ radial profiles from the \fable, \flamingo, \bahamas\ and \simba\ suites to the DESI Y1 + ACT BGS and LRG measurements \citep{Ried2025}.  As outlined in Section~\ref{subsubsec:sample_selectionggl}, galaxy-galaxy lensing-informed sample selection ensures that simulated samples are matched in mean halo mass to the observations. The mean halo mass estimates that we adopt follow \citet{siegel2025b}: $\log_{10}(\langle\ M_{500}[\mathrm{M_{\odot}}]\rangle)=13.31$ for the BGS sample and $\log_{10}(\langle\ M_{500}[\mathrm{M_{\odot}}]\rangle)=13.30$ for the LRG sample. For the LRG sample, we use the $z=0.75$ snapshot for simulation comparisons, matching the mean redshift of the data, and for BGS galaxies ($z=0.3$), we  use the nearest snapshots spanning $0.25<z<0.31$. Appendix~\ref{app:z_dep} shows over the redshift ranges $0.25<z<0.31$ and $0.65<z<0.86$, the kSZ profile evolves mildly, particularly in the inner radii, justifying our approach.

Table~\ref{tab:sim_ksz_stats_ggl} reports the properties of the simulated galaxy samples: stellar-mass thresholds, stacked counts, RMS velocities for cosmic variance corrections, and satellite fractions, as well as a quantification of the consistency between the simulated and observed kSZ profiles.  The goodness of fit between the simulations and observations is computed using
the full covariance matrices of the measurements, accounting for the strong correlations between the bins at large radii.  The minimum stellar-mass thresholds required to match the mean halo mass differ by up to 0.5 dex across simulations, reflecting differences in their stellar–halo mass relations.

Figure~\ref{fig:tksz_ggl_main} shows simulation–observation comparisons for the DESI+ACT BGS and LRG full-samples. With the exception of \fable\ and \xfable, all simulations lie within $2\sigma$ of the BGS data. In contrast, the fiducial simulations (\bahamas, \fable\ (and \xfable), and \flamingo) are $>10\sigma$ discrepant with the LRG measurement.  

Strong feedback variants (\flamingo\ $f_{\mathrm{gas}}-8\sigma$, \flamingo\ jets $f_{\mathrm{gas}}-4\sigma$, and \bahamas\ 8.0) significantly improve the fit relative to their counterparts, as does the \simba\ simulation, which despite having gas fractions similar to the fiducial \flamingo\ and \bahamas\ runs, produces a more extended gas distribution (Figure~\ref{fig:simprops}). These models also yield stronger matter power suppression, consistent with findings from earlier kSZ-based studies \citep{Bigwood2024, McCarthy2024,Hadzhiyska2024, Ried2025, Hadzhiyska2025, kovac2025,siegel2025b}. As shown in \citet{siegel2025b} and \citet{kovac2025}, this tension between fiducial simulations and kSZ is eased if the more gas-depleted fractions from eROSITA are adopted. Nevertheless, with the exception of the \flamingo\ $-8\sigma$ model, all simulations remain excluded at $>4\sigma$ against the LRG full-sample measurement. Because this discrepancy may be driven by the higher-mass LRG M3 and M4 bins (see Section~\ref{sec:actdata}, which are not yet fully understood), we focus on BGS, LRG M1, and LRG M2 in the quantitative comparisons that follow.

\subsection{Simulated \& observed mass dependence}\label{subsec:stellarmass_dep}

Figure~\ref{fig:tksz_massdep} (left) shows the mass dependence of the kSZ profiles at $z=0.75$, comparing simulations to the LRG M1 and M2 measurements. As before, we use halo masses from \citet{siegel2025b}:  $\log_{10}(\langle\ M_{500}[\mathrm{M_{\odot}}]\rangle)=12.91, 13.16$ and galaxy properties are reported in Table~\ref{tab:sim_ksz_stats_ggl}. Trends are intuitive: lower mass selections contain more galaxies, and satellite fractions decline from LRG M1 to LRG M2, consistent with expectations that massive satellites occupy the largest halos. Consistent with the BGS comparison, the strong-feedback variants of \flamingo\ and \bahamas, as well as \simba, reproduce the observed profiles within $<2\sigma$ when BGS, M1, and M2 are combined (Table~\ref{tab:sigma}). By contrast, fiducial models are excluded at $>3\sigma$.

The right panel of Figure~\ref{fig:tksz_massdep} highlights the kSZ amplitude measured at $\theta = 2.25$~arcmin across all simulations, including the LRG M3 and M4 measurements, and the photometric LRG M3 measurement \citep{Hadzhiyska2024}.  \simba, \bahamas\ 8.0, \flamingo\ $-8\sigma$, and \flamingo\ jets $f_{\mathrm{gas}}-4\sigma$  best reproduce the observed mass trend. However, none of the simulations reproduce the shallow halo mass dependence of the kSZ amplitude in the higher mass spectroscopic bins at $\log_{10}(M_{500}/\mathrm{M_{\odot}}) > 13.4$. In contrast, the strong feedback models and \simba\ better match the photometric LRG M3 measurement. 

Altogether, the simulations with the strongest matter power suppression provide the closest match to the ACT+DESI data. This requires either halo gas fractions lower than those implied by the X-ray calibration of fiducial models, closer to the eROSITA measurements \citep{Popesso2024,siegel2025b}, or AGN feedback implementations that produce more extended gas distributions, as in \simba.  \citet{Bigwood2025} presented a simulation suite encompassing a wide range of subgrid AGN feedback models beyond \fable\ and \xfable\  (which were both calibrated to pre-eROSITA halo gas mass fractions).  In light of recent evidence for lower halo gas fractions, it is now timely to revisit other models from the suite that were previously ruled out for producing overly gas-depleted clusters but remain consistent with galaxy observations, in order to better understand the physical mechanisms driving feedback.  The \simba\ simulation result demonstrates that it is possible to construct a feedback model that matches observed galaxy properties (e.g. the galaxy stellar mass function and the star formation rate density evolution) while providing a power spectrum suppression that matches that inferred from kSZ observations \citep{Bigwood2024, kovac2025}.  The requirement is that group-sized halos are highly evacuated and the dispersal of baryons is widespread \citep{borrow2020,sorini2022}.  This adds to the growing evidence that AGN feedback strongly impacts diffuse cosmic gas at low redshifts \citep[e.g.,][]{Christiansen2020, Tillman2023, vandaalen:2020}.  

\subsection{kSZ in a hypercube}\label{sec:antillescompare}
\begin{figure*}
\centering
\includegraphics[width=0.99\textwidth]
{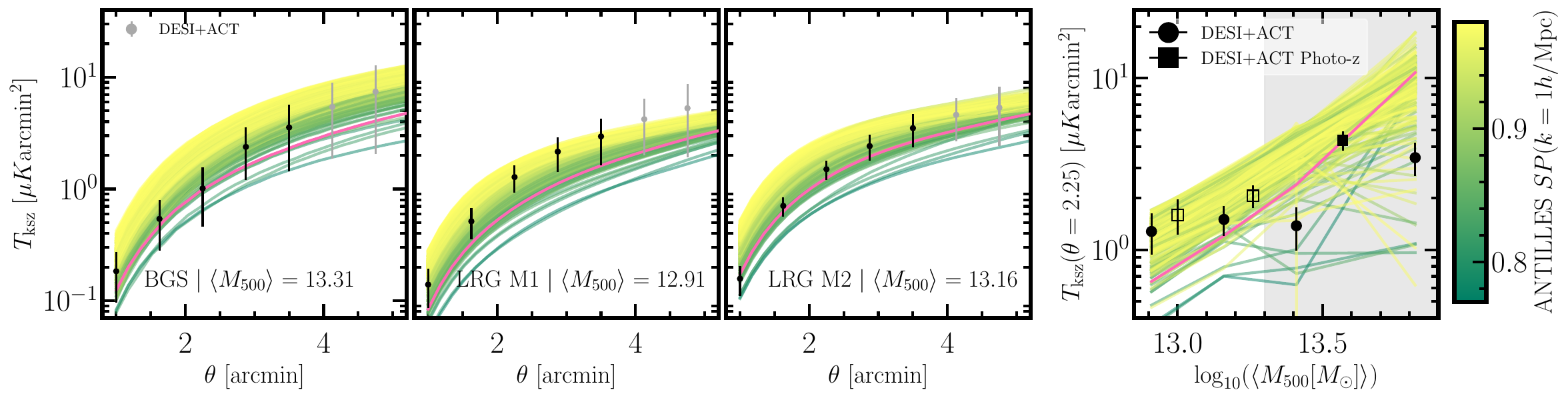}
    \caption{\textit{Left}: The stacked radial kSZ profile in \antilles\ in comparison to the $z\approx0.3$ DESI Y1 + ACT BGS measurements and $z\approx0.75$ LRG measurements in two stellar mass bins \citep[black data points;][]{Ried2025}.   We measure the kSZ signal in simulations within a CAP filter centered on galaxies, with galaxy samples constructed using a stellar mass–based selection, ensuring that the mean halo masses, $\langle \log_{10}(M_{\rm 500}/\mathrm{M_{\odot}}) \rangle$, in each bin matched those estimated from galaxy–galaxy lensing measurements reported in \citet{siegel2025b}.   The \antilles\ simulations are colour-coded by the suppression of the matter power spectrum due to baryonic effects at $k=1~\mathrm{Mpc}/h,\ z=0$.  \textit{Right:} The mass dependence of the amplitude of the kSZ 
    profile at $z=0.75$, measured at $\theta=2.25~\mathrm{arcmin}$.  We compare to the spectroscopic DESI Y1 + ACT measurements \citep{Ried2025} and the photometric DESI Y1 + ACT LRG M1, M2 and M3 bins \citep{Hadzhiyska2024}.  Since the photometric DESI Y1 + ACT LRG M1 and M2 bins follow the mass trend of the spectroscopic measurements, we do not make measurements for these bins, and we show these data points unfilled.  We shade the region of the plot $\log_{10}(M_{500}/\mathrm{M_{\odot}})>13.3$, because of the different mass trend of the kSZ amplitude exhibited between the spectroscopic and photometric measurements in this mass range.  The pink line represents the \antilles\ box that provides the best combined fit to the DESI + ACT BGS, LRG M1 and LRG M2 kSZ measurements.}
    \label{fig:antilles_ksz}
\end{figure*}

The \antilles\ hypercube of 400 $(100~\mathrm{Mpc}/h)^3$ simulations spans a wide range of feedback parameter space, generating large variation in matter power suppression, gas fractions, and gas density profiles. Here we compare the suite to the DESI+ACT BGS and LRG kSZ measurements to search for compatible models.

The left panel of Figure~\ref{fig:antilles_ksz} shows that  \antilles\ broadly spans the observational data. However, the simulated profiles are generally shallower than the LRG observations. Within the suite, the best combined fit to the BGS, LRG M1, and LRG M2 stacks is obtained for the model using the \textsc{anarchy} hydrodynamics scheme with parameters $v_w[\mathrm{km,s^{-1}}]=265.15$, $\eta_w=2.82$, $\log_{10}(T_{\mathrm{heat}}[\mathrm{K}])=8.45$, $n_{\mathrm{heat}}=19$, and $\log_{10}(n^*_{\mathrm{H,BH}}[\mathrm{cm}^{-3}])=-1.24$. This model is consistent with the data within $1.29\sigma$ and is shown as the pink line in Figures~\ref{fig:antilles_ksz} and \ref{fig:simprops}. Notably, it represents one of the most extreme feedback models in the suite, with matter power suppression and depleted baryon fractions exceeding even the strongest \bahamas\ and \flamingo\ variants.

The right panel of Figure~\ref{fig:antilles_ksz} compares the mass dependence of the kSZ amplitude measured at $\theta = 2.25$~arcmin in \antilles\ to the DESI + ACT LRG stellar mass binned measurements, and the analogous photometric measurements.  Consistent with our simulation study in Section~\ref{subsec:stellarmass_dep}, the suite fails to reproduce the spectroscopic mass dependence at $M_{500} \geq 2\times 10^{13} \mathrm{M_{\odot}}$.  

In summary, the \antilles\ hypercube successfully spans the DESI+ACT measurements and yields a formal best-fitting model consistent with the combined BGS, LRG M1, and LRG M2 constraints.  However, this preferred model is one of the most extreme feedback prescriptions in the suite and fails to reproduce the observed halo-mass dependence of the spectroscoic measurements, alike the other simulations we study.


\section{The kSZ effect as an indicator for the suppression of the matter power spectrum} \label{sec:kszsummary}



\begin{figure*}
\centering

\includegraphics[width=0.9\textwidth,keepaspectratio]{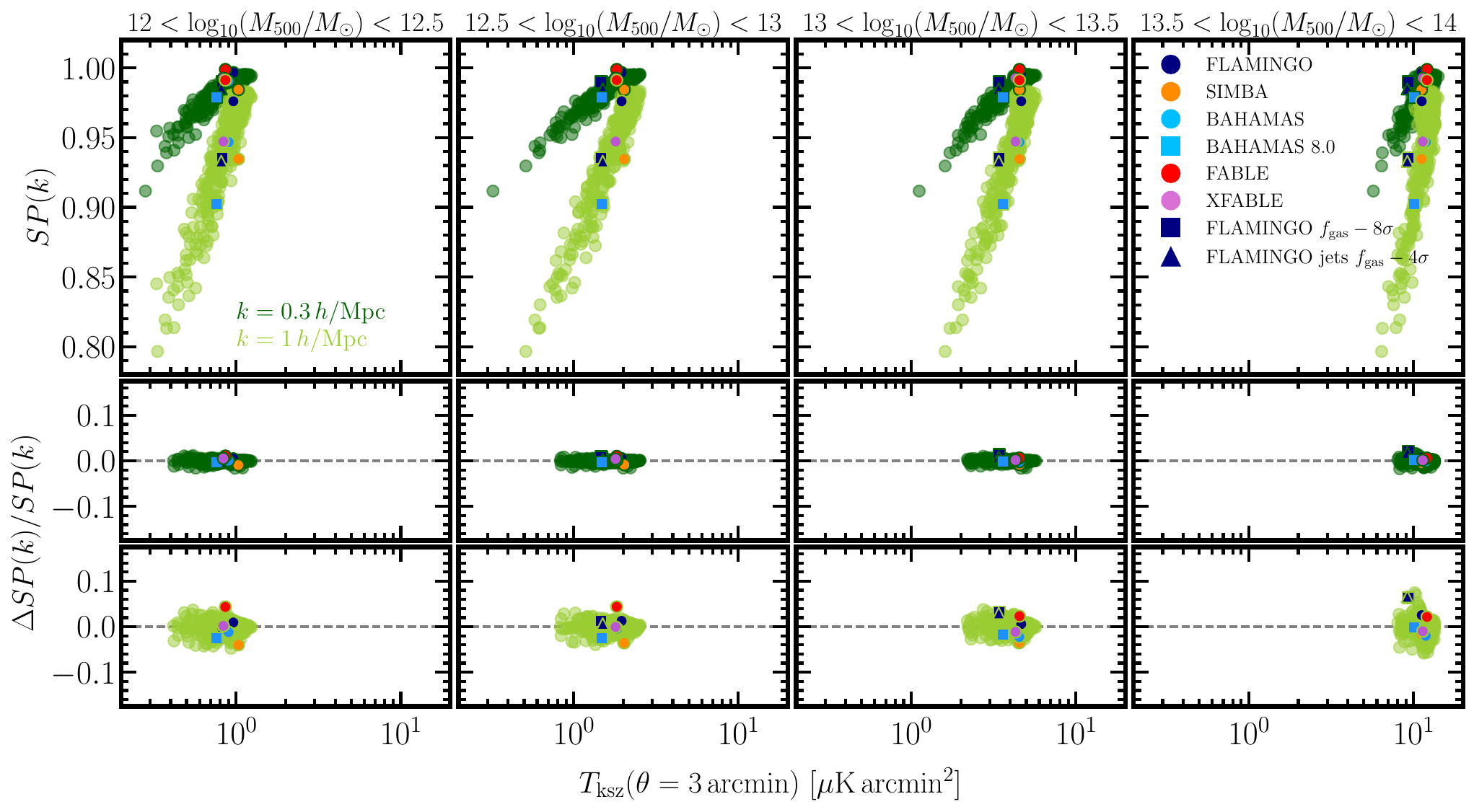}
\includegraphics[width=0.9\textwidth,keepaspectratio]{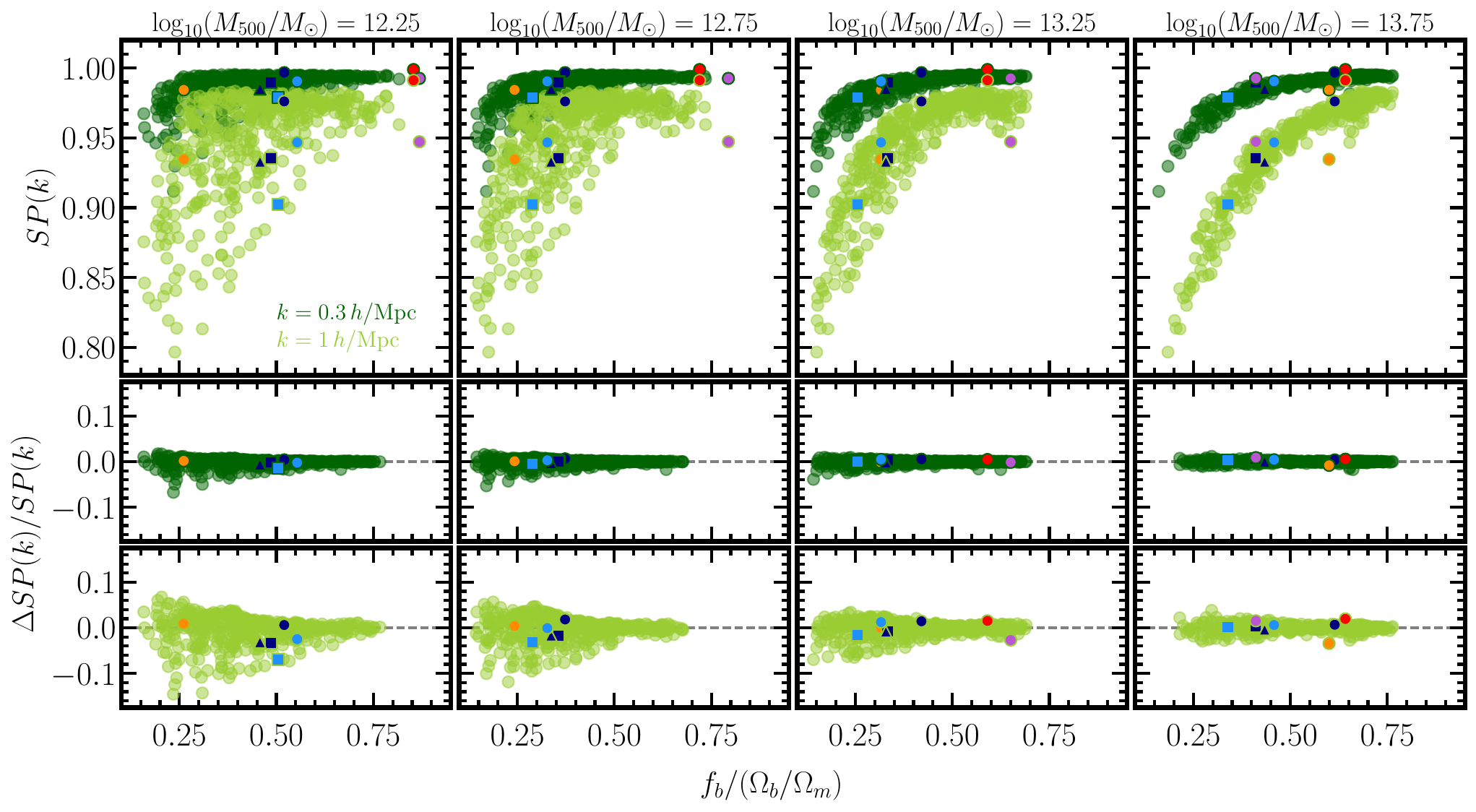}
    \caption{\textit{Upper plot: }the relationship between the $z=0.75$ kSZ amplitude measured at 3~arcmin, $T_{\mathrm{ksz}}(\theta=3~\mathrm{arcmin})$ and the suppression of the matter power spectrum, $SP(k)$ at $z=0$.   In all panels we plot the 400 \antilles\ boxes (light and dark green), as well as \fable\ (red), \xfable\ (purple), \flamingo\ (navy solid), \flamingo$f_{\mathrm{gas}}-8\sigma$ (navy dashed), \flamingo\ jets $f_{\mathrm{gas}}-4\sigma$ (navy dashed), \bahamas\ (blue solid), \bahamas\ 8.0 (blue dashed) and \simba\ (orange). The four top panels show $T_{\mathrm{ksz}}(\theta=3~\mathrm{arcmin})$ measured for objects stacked in four halo mass bins, following Figure~\ref{fig:halo_mass_dep}.  \textit{Lower plot:} the relationship between the $z=0$ median baryon fraction measured within $r_{500}$, $f_{\rm b}/(\Omega_{\rm b}/\Omega_{\rm m})$, and the suppression of the matter power spectrum, $SP(k)$, shown for the simulations as above.  The four top panels show $f_{\rm b}/(\Omega_{\rm b}/\Omega_{\rm m})$ measured at four halo masses.  In both plots, each panel shows $SP(k)$ measured at $k=0.3h/\mathrm{Mpc}$ (plotted as a dark green filled circle for \antilles\ and other colors for other simulations) and $k=1h/\mathrm{Mpc}$ (plotted as a light green filled circle for \antilles\ and other colors for other simulations).  The lower two sub-panels in each plot show the fractional scatter of the simulation points relative to the median relation, $\Delta SP(k)/SP(k)$.}
    \label{fig:antilles_tksz_corr}
\end{figure*}

Modeling the non-linear matter distribution requires understanding how baryonic feedback suppresses the power spectrum. One established probe is the baryon fraction, $f_{\rm b}/(\Omega_{\rm b}/\Omega_{\rm m})$, measured within $r500$, which correlates exponentially with suppression at cluster scales \citep{vanDaalen2011,vandaalen:2020,Salcido2023,Schaller2024}. The kSZ effect offers a complementary approach: it probes ionized gas out to several $r500$, accesses lower halo masses through its linear dependence on electron density, and therefore may provide a more powerful predictor of the suppression of the matter power spectrum.  In this section, we use the \antilles\ hypercube to test whether the kSZ effect can serve as a predictor of baryonic suppression of the matter power spectrum and quantify this relation, contrasting it with the traditional indicator of baryon fractions \citep{vandaalen:2020, Semboloni2011,Sembolini2013}. We find three main results:
\begin{itemize}
    \item The kSZ amplitude correlates tightly with the suppression of the matter power spectrum, especially at $k=1 h/\mathrm{Mpc}$
    \item This correlation remains robust down to lower halo masses, unlike halo baryon fraction measurements, for which the correlation with the suppression degrades outside the cluster regime.
    \item The kSZ signal therefore provides complementary, and in some cases superior, predictive power compared to $f_{\rm b}$.
\end{itemize}

The upper plot of Figure~\ref{fig:antilles_tksz_corr} shows the relation between the kSZ amplitude, $T_{\mathrm{ksz}}(\theta=3~\mathrm{arcmin})$ and the suppression of the matter power spectrum, $SP(k)$ at $z=0$. Across the 400 \antilles\ simulations (light and dark green), and others (\fable, \xfable, \flamingo, \flamingo$f_{\mathrm{gas}}-8\sigma$, \flamingo\ jets $f_{\mathrm{gas}}-4\sigma$, \bahamas, \bahamas\ 8.0  and \simba), a clear gradient emerges for $T_{\mathrm{ksz}}(SP(k))$ at $k=1h/\mathrm{Mpc}$, where suppression varies most strongly, and also at larger scales, at  $k=0.3h/\mathrm{Mpc}$. Residuals show scatter about the median relation below $\sim5$\% (middle and bottom rows), demonstrating that the kSZ amplitude is a sensitive predictor of suppression at non-linear scales. 

The lower plot shows the analogous relation for baryon fractions\footnote{We measure the baryon fraction as the sum of gas and stellar mass within $r_{500}$, and do not exclude the cold gas fraction, since the impact of feedback on the matter power spectrum is due to gravitational processes and therefore on the total mass content.}. As found previously \citep{vandaalen:2020,Salcido2023}, $f_{\rm b}$ correlates cleanly with suppression at cluster masses, with scatter comparable to kSZ ($\sim5$\%). However, at lower masses the relation degrades rapidly, with much larger scatter than seen for the kSZ. Gas fractions alone (Appendix~\ref{fig:fgas=pk}) perform better than total baryon fractions at group scales, helping explain why the kSZ, sensitive only to ionized gas, retains a cleaner correlation.

A key distinction is that the predictive power of the kSZ extends to lower halo masses. The correlation between $T_{\mathrm{ksz}}(\theta=3~\mathrm{arcmin})$ and $SP(k)$ remains tight down to $M_{500}\approx 10^{12}\mathrm{M_{\odot}}$. This stability may reflect two factors: (i) the kSZ probes extended gas distributions beyond $r_{500}$, and (ii) stacking halos across a mass range aggregates information that sharpens the relation. In contrast, halo baryon fraction measurements degrade once halos fall below cluster mass at the $k$-scales shown.  However, baryon fractions in lower-mass halos are correlated more strongly with the matter power spectrum suppression on smaller scales than those considered here, and combining measurements across a range of apertures and halo masses can enable reconstruction of the suppression \citep{Salcido2023, salcido2025,vanDaalen2025}.

Taken together, these results show that the kSZ effect is not only consistent with baryon fraction methods but also provides additional advantages. It correlates more strongly with suppression at $k=1h/\mathrm{Mpc}$, remains robust at lower masses, and directly probes the ionized gas driving feedback effects. This establishes the kSZ as a powerful new benchmark observable for modeling baryonic suppression in cosmological analyses. As with $f_{\rm b}$, a next step is to calibrate a quantitative relation between kSZ amplitude and suppression, enabling it to play the same role in constraining small-scale cosmology.

\section{Conclusions}



Cosmological hydrodynamical simulations can reproduce a wide range of galaxy properties, but diverge in their predictions for the fraction of gas expelled from halos, the radial extent of that gas, and the resulting impact on the large-scale matter distribution (Figure~\ref{fig:simprops}).
In this work, we establish the stacked kSZ effect profile as a new benchmark for feedback models and present a roadmap for robust simulation–data comparison.
We analyze eight simulations spanning diverse feedback prescriptions, \fable, \xfable, \simba, \bahamas\ and \bahamas-8.0 and \flamingo\ variants, and compare their predictions with the latest DESI+ACT kSZ observations \citep{Ried2025}. Our key takeaways are: 
\begin{itemize}
\item \textbf{Robust like-for-like comparison between the simulations and observations}: To achieve this, it is critical to consider the halo mass and satellite fraction in the sample selection, cosmic variance and mis-centering effects:

\begin{itemize}
    \item We select a simulated galaxy sample by stellar mass with a mean halo mass that matches the observed sample by jointly fitting to the galaxy-galaxy lensing measurements (Section~\ref{subsubsec:sample_selectionggl}), following \citet{McCarthy2024, siegel2025b}. 
    \item We effectively mitigate cosmic variance (particularly for simulation volumes $< 200\mathrm{Mpc}/h$), by correcting the RMS of radial galaxy velocities to match that of the larger-volume \flamingo\ simulation, and by quantifying the kSZ amplitude offsets that arise from different line-of-sight projections (Section~\ref{sec:cosmicvariance}).
    \item Because we use self-consistent cosmological hydrodynamical simulations, both satellite galaxies and miscentering effects are naturally incorporated through the sample selection used for comparison with the data.  While there remains uncertainty in the satellite fractions of the observed DESI samples, our simulation samples results are broadly consistent with the literature. 
\end{itemize}
\item \textbf{Comparison to DESI + ACT:}
Of the simulations that we study, the fiducial models are in tension ($>3\sigma$) with DESI + ACT BGS and stellar mass binned LRG kSZ measurements, with the exception of SIMBA (Figures~\ref{fig:tksz_ggl_main}, \ref{fig:tksz_massdep} and  Tables~\ref{tab:sim_ksz_stats_ggl}, \ref{tab:sigma}). Strong-feedback variants, such as \bahamas\ 8.0, \flamingo\ $f_{\rm gas}-8\sigma$ and jets $f_{\rm gas}-4\sigma$, provide markedly better agreement. Therefore, simulations producing the strongest suppression of the matter power spectrum best match the data, either requiring lower cluster gas fractions than the X-ray observations used to calibrate the simulations (consistent with recent eROSITA measurements) or a feedback mechanism that produces a more extended gas distribution.

\item \textbf{Halo mass dependence:} None of the simulations we study reproduce the halo mass dependence observed in the DESI+ACT spectroscopic LRG measurements at $\log_{10}(M_{500}/\mathrm{M_{\odot}}) > 13.4$ \citep{Ried2025}. By contrast, the strong feedback simulation variants more closely follow the trend seen in the DESI+ACT photometric LRG kSZ measurements \citep{Hadzhiyska2024}. 

\item \textbf{Predictive power:} Within the \antilles\ hypercube and across the other simulations we study, the amplitude of the kSZ effect, $T_{\mathrm{ksz}}(\theta=3~\mathrm{arcmin})$ is highly correlated with the $z=0$ suppression of the matter power spectrum, $SP(k=1h/\mathrm{Mpc})$ (Figure~\ref{fig:antilles_tksz_corr}). For high mass halos ($13.5 < \log_{10}(M_{500}[\mathrm{M_{\odot}}]) < 14$), the scatter of simulations around the binned median relation is comparable to the well-studied relationship with the baryon fraction in groups and clusters $f_{\rm b}$ and $T_{\mathrm{ksz}}(\theta=3~\mathrm{arcmin})$ with $SP(k)$ \citep{vandaalen:2020, Salcido2023}. However, at lower masses ($12 < M_{500}[\mathrm{M_{\odot}}] < 13.5$), the correlation between $T_{\mathrm{ksz}}$ and $SP(k)$ is significantly improved (with the scatter around the median relation remaining $\lesssim5\%$).  Furthermore, $T_{\mathrm{ksz}}(\theta=3~\mathrm{arcmin})$ shows a clearer correlation with the larger-scale suppression of the matter power spectrum measured at $SP(k=0.3h/\mathrm{Mpc})$ than is the case for baryon fraction measurements.  This speaks to the potential of using the kSZ effect as a predictor for the suppression of the matter power spectrum and motivates future work to establish a more quantitative relationship. 
\end{itemize}

These findings highlight potential shortcomings in the calibration of galaxy formation models of cosmological hydrodynamical simulations, providing quantitative evidence that current fiducial simulations may underestimate the impact of baryonic feedback on both the gas distribution and the suppression of the matter power spectrum. Furthermore, our results demonstrate the potential of the kSZ effect to constrain feedback physics in regimes currently inaccessible to X-ray measurements, particularly in lower-mass halos and at extended radii.  
Looking ahead, upcoming DESI data releases, together with high-resolution CMB maps from ACT and the Simons Observatory \citep{simons2019}, are forecasted to improve the kSZ signal-to-noise ratio by more than an order of magnitude \citep{battaglia2019}. This will allow precision measurements of kSZ profiles as a function of halo mass, redshift, and angular scale, paving the way toward a quantitative calibration of feedback models in cosmological simulations. Such progress is critical not only for advancing our understanding galaxy formation, but also for ensuring robust cosmological inference on the non-linear scales probed by Stage-IV weak lensing surveys, where the uncertainties in baryonic feedback modelling currently represent the dominant limiting systematic \citep{DESKIDS2023}.  Ultimately, in this paper we have laid the groundwork for a robust interpretation of feedback using the kSZ effect.  In combination with the upcoming measurements, this promises to further our understanding of how galaxies regulate their baryons and reduce feedback modeling uncertainties, providing a path toward precision cosmology on small scales as well as powerful constraints on the circumgalactic gas profiles of massive galaxies.

\section*{Acknowledgements}
We thank Bernardita Ried Guachalla for helpful feedback and making the data available to us, and Debora Sijacki for useful discussions and guidance in using the FABLE simulations.  We also thank Weiguang Cui for their guidance in using SIMBA.  Leah Bigwood acknowledges support from the Science and Technology Facilities Council (STFC). I. McCarthy acknowledges support from the STFC (grant number ST/Y002733/1). This project has received funding from the European Research Council (ERC) under the European Union’s Horizon 2020 research and innovation programme (grant agreement No 769130). This work used the DiRAC@Durham
facility managed by the Institute for Computational
Cosmology on behalf of the STFC Distributed Research Utilizing Advanced Computing (DiRAC) High Performance
Computing Facility (www.dirac.ac.uk). The equipment
was funded by BEIS capital funding via STFC capital
grants ST/K00042X/1, ST/P002293/1, ST/R002371/1,
and ST/S002502/1, Durham University and STFC operations grant ST/R000832/1. DiRAC is part of the
National e-Infrastructure.

\section*{Data Availability}
The simulation kSZ profiles measured in this work are available from the corresponding author upon reasonable request. The \simba\ and \bahamas\ simulations are publicly available, with \fable\ also soon to be made public.  The \flamingo\ simulation data will eventually be made publicly available, though
we note that the data volume (several petabytes) may prohibit us
from simply placing the raw data on a server. In the meantime,
people interested in using the simulations are encouraged to contact the core \flamingo\ team.



\bibliographystyle{mnras}
\bibliography{references} 




\appendix

\section{Mean halo masses of abundance-matched samples}

\begin{figure}
\centering
\includegraphics[width=0.5\textwidth]{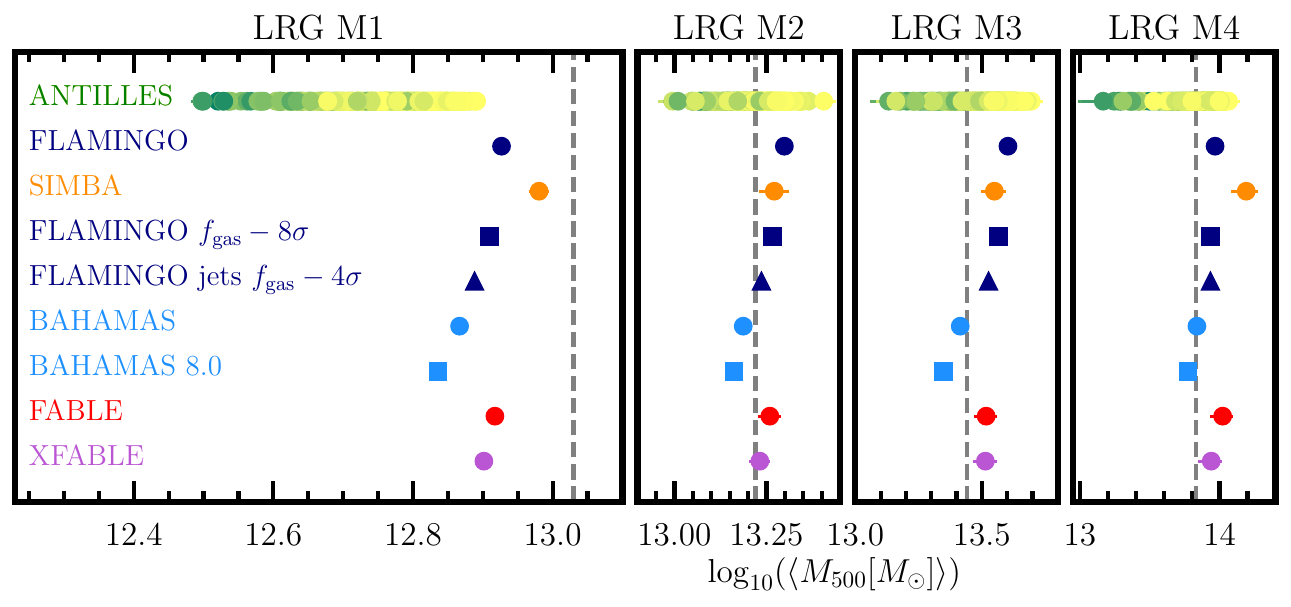}
    \caption{ The mean halo masses, $M_{500}$, of simulated objects in the four $M_{*}$ bins reported in \citet{Ried2025}; $\log_{10}(M_*/\mathrm{M_{\odot}}) \in \mathrm{LRG\ M1:}[10.5,11.2], \mathrm{LRG\ M2:}[11.2,11.4],\mathrm{LRG\ M3:}[11.4,11.6], $ $\mathrm{LRG\ M4:}[11.6,12.5]$, shown for each of the simulations we study.  In the simulations we reassign stellar masses by abundance matching to the \citet{Behroozi2019} galaxy stellar mass function, defining $M_{*}$ for each galaxy within a 50 pkpc spherical aperture centered on the subhalo particle with the minimum gravitational potential.  We plot the 400 \antilles\ boxes, as well as \fable\, \xfable\, \flamingo\, \flamingo$f_{\mathrm{gas}}-8\sigma$, \flamingo\ jets $f_{\mathrm{gas}}-4\sigma$, \bahamas\, \bahamas\ 8.0 and \simba\.  The \antilles\ simulations are colour-coded by the suppression of the matter power spectrum due to baryonic effects measured at $k=1~\mathrm{Mpc}/h$.  The grey dashed line shows the mean halo mass for each stellar mass bin estimated in \citet{siegel2025b}, attained via constructing galaxy samples in the fiducial \flamingo\ box to match the observed galaxy galaxy lensing profiles for the four DESI Y1 + ACT stellar mss bins (see Section~\ref{subsubsec:sample_selectionggl}). 
    }
    \label{fig:m500_mstar_selected}
\end{figure}

In Figure~\ref{fig:m500_mstar_selected} we show the mean halo masses of galaxy samples obtained through abundance matching for the LRG M1-M4 stellar mass bins, across the simulations we study.

\section{Simulating other measurements of kSZ profiles}\label{app:highmassbins}

\begin{figure*}
\centering
\includegraphics[width=0.85\textwidth]{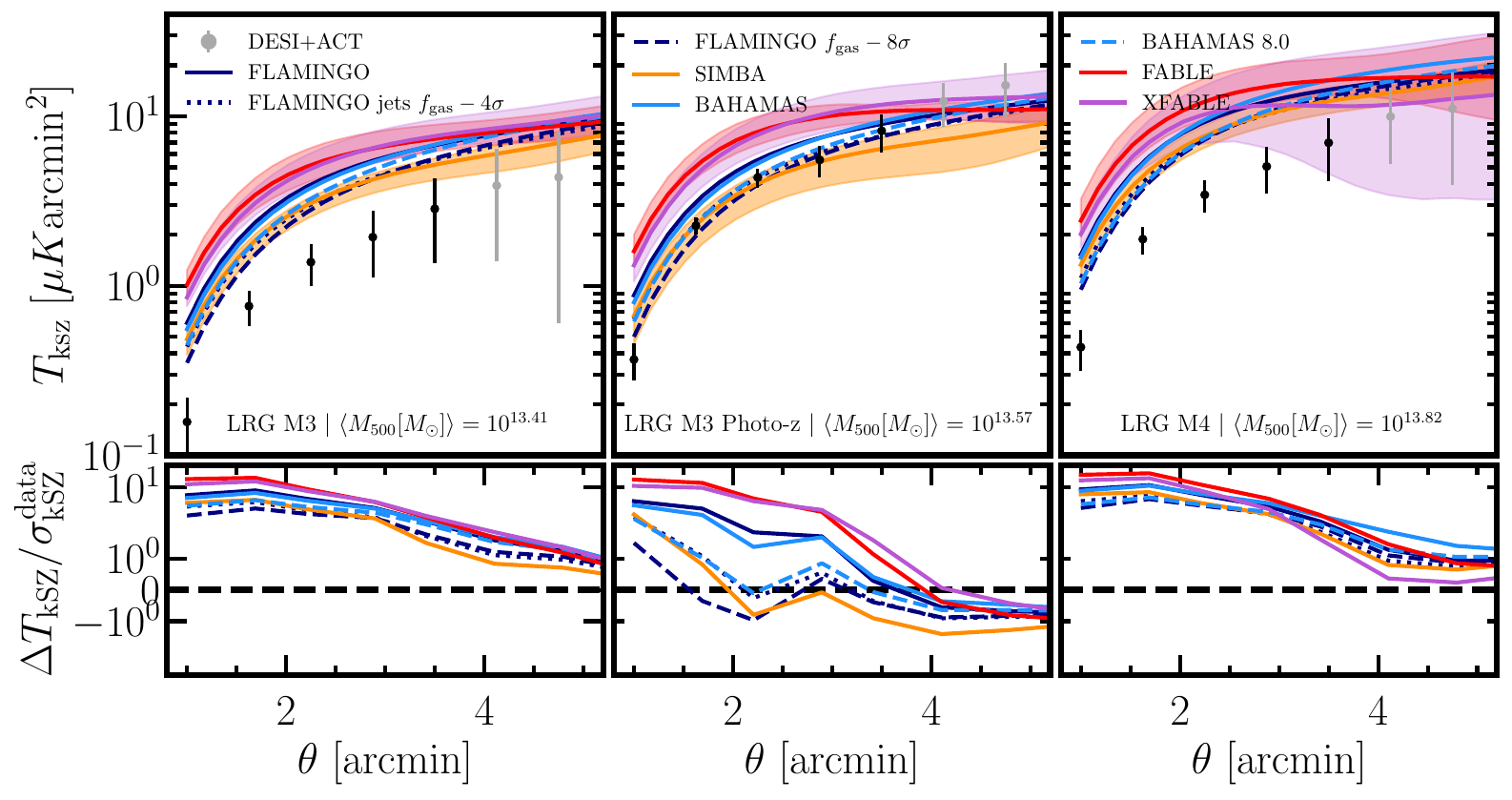}
\caption{As Figure~\ref{fig:tksz_ggl_main}, but comparing to the DESI Y1 + ACT LRG M3 ($\log_{10}(M_*/\mathrm{M_{\odot}}) \in [11.4,11.6]$) and M4 ($\log_{10}(M_*/\mathrm{M_{\odot}}) \in [11.6,12.5]$) stellar mass bins \citep[black data points;][]{Ried2025}, and the DESI Y1 + ACT photometric $z\approx0.75$ LRG M3 ($\log_{10}(M_*/\mathrm{M_{\odot}}) \in [11.5,12.0]$) measurements. }
    \label{fig:highmass}
\end{figure*}

In Figure~\ref{fig:highmass} we plot a simulation comparison to the DESI + ACT LRG M3 ($\log_{10}(M_*/\mathrm{M_{\odot}}) \in [11.4,11.6]$) and M4 ($\log_{10}(M_*/\mathrm{M_{\odot}}) \in [11.6,12.5]$) stellar mass binned kSZ measurements, as well as the DESI Y1 photometric LRG M3 ($\log_{10}(M_*/\mathrm{M_{\odot}}) \in [11.5,12.0]$) measurement presented in \citet{Hadzhiyska2024}.  We select samples to match the mean halo mass estimates of $\log_{10}(\langle\ M_{500}[\mathrm{M_{\odot}}]\rangle)=13.41,13.57,13.81$ reported by \citet{siegel2025b} for LRG M3, LRG M3 Photo-z (noting that the spectroscopic and photometric LRG M3 bins do not correspond to the same stellar-mass limits), and LRG M4 bins respectively.

Table~\ref{tab:sim_ksz_stats_ggl_highmass} presents the properties of the galaxy sample and fit to the data. Due to the small number statistics of galaxies in these higher-mass samples, we notice that the error region for the 100~Mpc/$h$ volumes, denoting the variance in the KSZ profile arising from projecting along the three different independent box axes, are larger than those for the lower-mass LRG M1 and M2 bins. Nevertheless, we find that \textit{all} simulations studied, including the strongest feedback variants, are significantly disfavored by the data.  The LRG M3 photometric bin, however, shows a picture more consistent with the LRG M1, M2 and BGS bins presented in the main text: while fiducial simulations are ruled out, the strong feedback simulation variants and \simba\ provide a reasonable fit.  This warrants further investigation into the observational data to understand the discrepancy between the photometric and spectroscopic measurements, before accurate insights into feedback can be drawn in this mass range.

\begin{table}
\centering
\caption{The combined number of standard deviations by which each simulation prediction deviates from the observations, computed using the full DESI + ACT BGS and LRG full sample measurements, $\sigma_{\rm BGS+LRG}$.} \label{tab:sigmahighmass}

\begin{tabular}{ccc}
\hline \hline
Simulation &  $\sigma_{\rm BGS+LRG}$\\
\hline
BAHAMAS &  $>$10  \\
BAHAMAS 8.0  & 4.6 \\
FLAMINGO & $>$10 \\
FLAMINGO $f_{\rm gas} - 8\sigma$ &   3.0 \\
FLAMINGO jets $f_{\rm gas} - 4\sigma$ &  4.7 \\
SIMBA &  5.5  \\
FABLE &  $>$10\\
XFABLE &  $>$10 \\
\hline \hline
\end{tabular}

\end{table}

\begin{table*}
\centering
\caption{Properties of the galaxy samples selected in each simulation for a like-for-like comparison with the stacked kSZ signals from DESI Y1 + ACT \citep{Ried2025}.  We report the minimum stellar mass limit used to select the sample ($\log_{10}(\langle M_{*, \mathrm{min}}\rangle)$), the mean halo mass  ($\log_{10}(\langle M_{500}[\mathrm{M_{\odot}}]\rangle)$), the root-mean square of line-of-sight velocities (${\langle v_{\mathrm{LOS}}^2 \rangle}^{\frac{1}{2}}$), the number of galaxies in the stack ($N_{\rm stack}$), the satellite fraction ($f_{\rm sat}$) and the number of standard deviations by which the simulations deviate from the observations ($\sigma$). For all the simulations except the \flamingo\ and \bahamas\ suites, $N_{\rm stack}$ is the sum of the number of halos used in all the three projections along the line-of-sight axis, and the other quantities are the average of the quantities over the three axes. \label{tab:sim_ksz_stats_ggl_highmass}}

\begin{tabular}{cccccccc}
\hline \hline
Bin & Simulation & $\log_{10}(\langle M_{*, \mathrm{min}}[\mathrm{M_{\odot}}]\rangle)$& $\log_{10}(\langle M_{500}[\mathrm{M_{\odot}}]\rangle)$ & ${\langle v_{\mathrm{LOS}}^2 \rangle}^{\frac{1}{2}}$ [km/s] & $N_{\rm stack}$ & $f_{\rm sat}$ [\%] & $\sigma$\\
\hline
\multirow{9}{*}{LRG M3} 
  & BAHAMAS & 11.33 & 13.41 & 270.84 & 7777 & 13.6 &  $>10$ \\
  & BAHAMAS 8.0 & 11.35 & 13.42 & 266.67 & 5182 & 13.3 & 6.00 \\
  & FLAMINGO & 11.29 & 13.41 & 298.53 & 76669 & 17.8 &  $>10$ \\
  & FLAMINGO $f_{\rm gas} - 8\sigma$ & 11.32 & 13.41 & 298.75 & 64065 & 17.3 & 4.03 \\
  & FLAMINGO jets $f_{\rm gas} - 4\sigma$ & 11.23 & 13.41 & 297.60 & 53923 & 16.8 & 5.37 \\
  & SIMBA & 11.20 & 13.42 & 198.01 & 743 & 14.2 & 6.13 \\
  & FABLE & 11.24 & 13.43 & 228.90 & 663 & 12.1 & $>10$ \\
  & XFABLE & 11.34 & 13.44 & 234.61 & 471 & 13.6 & $>10$ \\
\cline{1-8}
\multirow{9}{*}{LRG M3 (photo-z)} 

  & BAHAMAS & 11.45 & 13.57 & 266.98 & 2848 & 11.0 & 5.00\\
  & BAHAMAS 8.0 & 11.45 & 13.58 & 269.71 & 1967 & 11.9 & 1.79 \\
  & FLAMINGO & 11.42 & 13.57 & 298.10 & 33021 & 14.3 & 6.31 \\
  & FLAMINGO $f_{\rm gas} - 8\sigma$ & 11.44 & 13.57 & 300.92 & 27630 & 14.1 & 0.85 \\
  & FLAMINGO jets $f_{\rm gas} - 4\sigma$ & 11.34 & 13.57 & 298.23 & 23860 & 13.9 & 1.96 \\
  & SIMBA & 11.36 & 13.58 & 212.31 & 334 & 12.3 & 2.42 \\
  & FABLE & 11.42 & 13.59 & 228.94 & 237 & 11.8 & $>10$\\
  & XFABLE & 11.47 & 13.60 & 221.07 & 186 & 15.1 & $>10$ \\
\cline{1-8}
\multirow{9}{*}{LRG M4} 

  & BAHAMAS & 11.57 & 13.83 & 256.63 & 750 & 10.1 & $>10$ \\
  & BAHAMAS 8.0 & 11.57 & 13.82 & 267.77 & 547 & 9.0 & 6.77 \\
  & FLAMINGO & 11.61 & 13.82 & 301.13 & 8700 & 9.6 & $>10$ \\
  & FLAMINGO $f_{\rm gas} - 8\sigma$ & 11.63 & 13.82 & 302.51 & 7046 & 9.0 & 6.37 \\
  & FLAMINGO jets $f_{\rm gas} - 4\sigma$ & 11.49 & 13.82 & 300.61 & 6621 & 9.2 & 7.43 \\
  & SIMBA & 11.57 & 13.84 & 204.95 & 112 & 10.4 & $>10$ \\
  & FABLE & 11.61 & 13.85 & 257.91 & 61 & 4.9 & $>10$ \\
  & XFABLE & 11.72 & 13.87 & 296.77 & 20 & 0.0 & $>10$ \\

\hline \hline
\end{tabular}
\end{table*}

\section{Properties of the halo mass binned galaxy selection}\label{app:details}
In Section~\ref{subsec:halomass_dep}, we presented the kSZ profile measured in each simulation in four $M_{500}$ bins: $\log_{10}(M_{500}/\mathrm{M_{\odot}}) \in [(12,12.5), (12.5,13),(13,13.5), (13.5,14)]$. Table~\ref{tab:sim_ksz_stats_halo} reports the properties of the galaxy samples in each simulation for each mass bin.

\begin{table*}
\centering
\caption{ Properties of the galaxy samples selected in each simulation for the four halo mass bins studied in Section~\ref{subsec:halomass_dep}: mean halo mass ($\log_{10}(\langle M_{500}[\mathrm{M_{\odot}}]\rangle)$), root-mean square of line-of-sight velocities (${\langle v_{\mathrm{LOS}}^2 \rangle}^{\frac{1}{2}}$), the number of galaxies to stack ($N_{\rm stack}$) for each halo mass bin in simulations.  For all the simulations except the \flamingo\ and \bahamas\ suites, $N_{\rm stack}$ is the sum of the number of halos used in all the three projections along the line-of-sight axis, and the other quantities are the average of the quantities over the three axes. 
\label{tab:sim_ksz_stats_halo}}

\begin{tabular}{ccccc}
\hline \hline
Halo mass bin [$\, M_\odot$] & Simulation & $\log_{10}(\langle M_{500}[\mathrm{M_{\odot}}]\rangle)$ & ${\langle v_{\mathrm{LOS}}^2 \rangle}^{\frac{1}{2}}$ [km/s] & $N_{\rm stack}$ \\
\hline

\multirow{9}{*}{$M_{500} = 10^{12} - 10^{12.5}$} 
  & ANTILLES & 12.21 - 12.23 & 196.47 - 206.40 & 3909 - 6291 \\
  & BAHAMAS & 12.22 & 263.27 & 136296 \\
  & BAHAMAS $\Theta_{\rm AGN}=8.0$ & 12.22 & 263.55  &  131717\\
  & FLAMINGO & 12.22 & 293.75 & 697857 \\
  & FLAMINGO $f_{\rm gas} - 8\sigma$ & 12.21 & 293.67 & 716622 \\
  & FLAMINGO jets $f_{\rm gas} - 4\sigma$ & 12.21 & 293.58 & 687642 \\
  & SIMBA & 12.21 & 206.07 & 5284 \\
  & FABLE & 12.23 & 214.20 & 6844 \\
  & XFABLE & 12.23 &213.73  &6821\\
\cline{2-5}
\multirow{9}{*}{$M_{500} = 10^{12.5} - 10^{13}$} 
  & ANTILLES & 12.71 - 12.73 & 198.25 - 213.56  & 1117 - 1728 \\
  & BAHAMAS & 12.72 & 263.86 & 33866 \\
  & BAHAMAS $\Theta_{\rm AGN}=8.0$ & 12.72 & 263.20 & 31936 \\
  & FLAMINGO & 12.72 & 292.42 & 173069 \\
  & FLAMINGO $f_{\rm gas} - 8\sigma$ & 12.72 & 292.30 & 168520 \\
  & FLAMINGO jets $f_{\rm gas} - 4\sigma$ & 12.72 & 292.69 & 163067 \\
  & SIMBA & 12.70 & 198.33 & 1295 \\
  & FABLE & 12.73  & 214.56 &  2025\\
  & XFABLE & 12.73 & 215.31  & 2077 \\
\cline{2-5}
\multirow{9}{*}{$M_{500} = 10^{13} - 10^{13.5}$} 
  & ANTILLES &13.19 - 13.25 & 168.71 - 185.07 & 245 -  425 \\
  & BAHAMAS & 13.21 & 264.15 & 8121 \\
  & BAHAMAS $\Theta_{\rm AGN}=8.0$ & 13.21 & 264.06 & 7636 \\
  & FLAMINGO & 13.21 & 291.06 & 42739 \\
  & FLAMINGO $f_{\rm gas} - 8\sigma$ & 13.21 & 291.74 & 40630 \\
  & FLAMINGO jets $f_{\rm gas} - 4\sigma$ & 13.21 & 290.83 & 39967 \\
  & SIMBA & 13.21 & 181.97 & 243 \\
  & FABLE & 13.19 & 219.05 & 472 \\
  & XFABLE & 13.18 & 217.78 & 478 \\
\cline{2-5}
\multirow{9}{*}{$M_{500} = 10^{13.5} - 10^{14}$} 
  & ANTILLES & 13.67 - 13.77 & 175.39 - 21645  & 41 - 89 \\
  & BAHAMAS & 13.71 & 269.02 & 1660 \\
  & BAHAMAS $\Theta_{\rm AGN}=8.0$ & 13.70 & 270.30 & 1530 \\
  & FLAMINGO & 13.70 & 291.07 & 9038 \\
  & FLAMINGO $f_{\rm gas} - 8\sigma$ & 13.70 & 291.13 & 8155 \\
  & FLAMINGO jets $f_{\rm gas} - 4\sigma$ & 13.70 & 291.01 & 8002 \\
  & SIMBA & 13.68 & 199.54 & 63 \\
  & FABLE & 13.69 & 223.41 & 83 \\
  & XFABLE & 13.67 & 224.96 & 76 \\

 
 
\hline \hline
\end{tabular}

\end{table*}

\section{The redshift evolution of the kSZ profile}\label{app:z_dep}

\begin{figure}
\centering
\includegraphics[width=0.85\columnwidth]{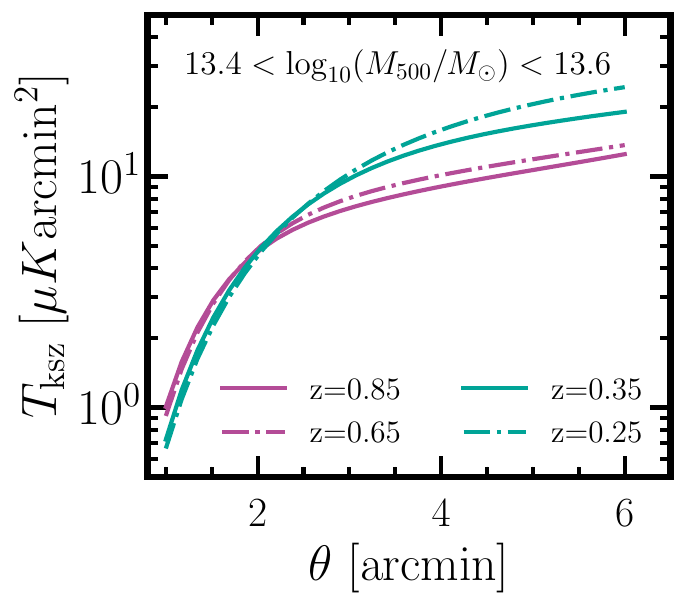}
\caption{The kSZ radial profile in the fiducial \flamingo\ box, measured by stacking galaxies with $13.4 < \log_{10}{M_{500}[\mathrm{M_{\odot}}]} < 13.6$ at redshifts $z=0.65, 0.85$ and $z=0.25, 0.35$ to demonstrate the impact of a slight mismatch in our choice of redshift in simulations and in the four DESI LRG and BGS bins.}
    \label{fig:ksz_zevol}
\end{figure}

\begin{figure}
\centering
\includegraphics[width=0.85\columnwidth]{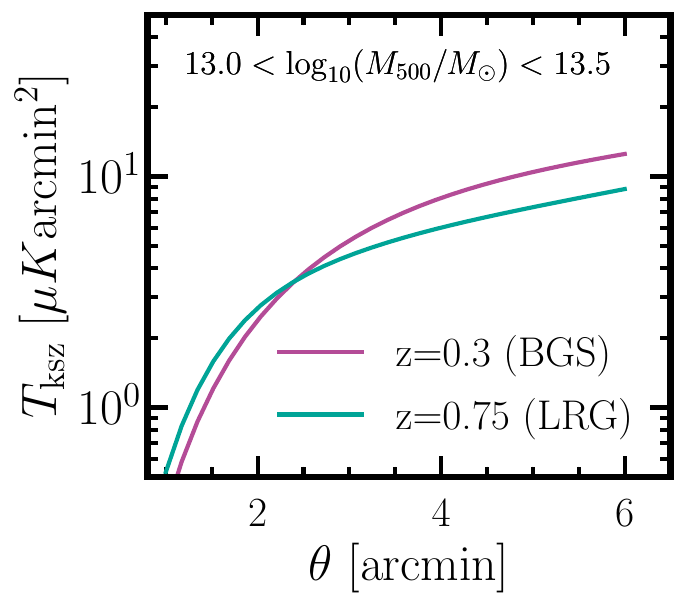}
\caption{The kSZ radial profile in the fiducial \flamingo\ box, measured by stacking galaxies with $13.0 < \log_{10}{M_{500}[\mathrm{M_{\odot}}]} < 13.5$ at redshifts $z=0.3$ (BGS sample-like redshift) and $0.75$ (LRG sample-like redshift).  }
    \label{fig:ksz_bgs_lrg}
\end{figure}

Described in Sec.~\ref{subsec:stellarmass_dep}, our kSZ profile measurement of the BGS galaxy sample in simulations used the snapshots and group catalogs at $0.25<z<0.31$, while the mean redshift of the observed BGS sample is $z=0.3$.  Similarly our kSZ profile measurement of the four LRG stellar mass binned samples in simulations (see Sec.~\ref{subsec:stellarmass_dep} and Appendix~\ref{app:highmassbins}) used the snapshots and group catalogs at $z=0.75$, while in \citet{Ried2025}, the mean redshift of their four stellar-mass selected galaxy samples ranges from $z=0.69$ to $z=0.76$. We do not attempt to match their redshifts perfectly for each mass bin due to the constraints on the available number of snapshots. Figure~\ref{fig:ksz_zevol} shows the kSZ profile of a galaxy sample selected in a relatively narrow $M_{500}$ bin of $13.4 < M_{500} < 13.6$ at BGS-like ($z=0.25$ and $z=0.35$) and LRG-like redshifts ($z=0.65$ and $z=0.85$) in the fiducial \flamingo\ box.  Given that the redshift mismatch has only a minor impact on the kSZ amplitude, particularly at small radii where the observed measurements are uncorrelated, we verify that our comparison to \citet{Ried2025} is robust to the redshift discrepancies in the simulations. 

Finally, for direct comparison with the third column of Figure~\ref{fig:halo_mass_dep}, we show in Figure~\ref{fig:ksz_bgs_lrg} the redshift evolution of the kSZ profile measured at BGS ($z=0.3$) and LRG-like ($z=0.75$) redshifts in the fiducial \flamingo\ box.

\section{kSZ profiles uncorrected for cosmic variance}\label{app:vlos}

\begin{figure*}
\centering
\includegraphics[width=0.95\textwidth]{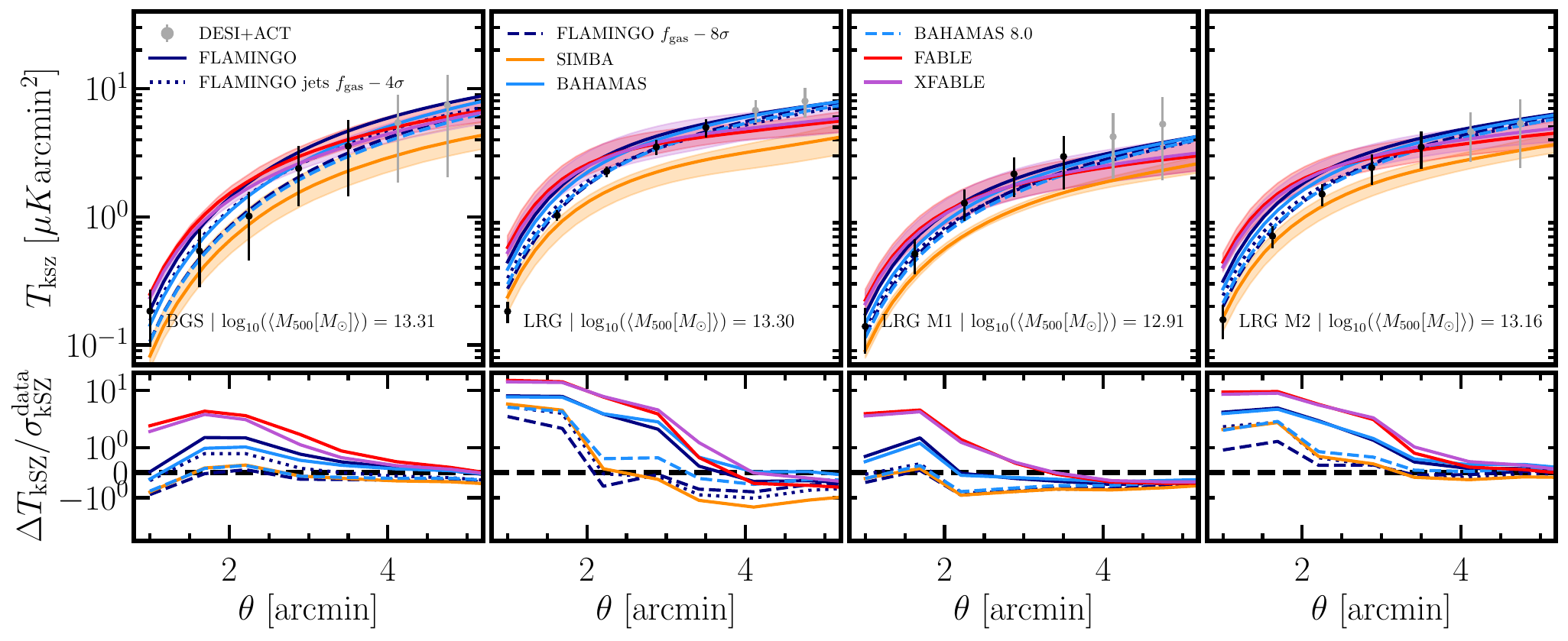}
\caption{Identical to Figure~\ref{fig:tksz_ggl_main} and Figure~\ref{fig:tksz_massdep}, but showing kSZ profiles without the correction for cosmic variance to the RMS of the radial velocities in the stack, as described in Section~\ref{sec:cosmicvariance}.}
    \label{fig:unscaled}
\end{figure*}

As described in Sec.~\ref{sec:cosmicvariance}, we found that the limited sample size of galaxies in small (e.g., 100 Mpc/$h$) simulation boxes can result in scatter in the amplitude of the measured kSZ profile. We found that this is predominantly driven by variations in the RMS of the halo peculiar velocities around the value one would predict given the simulation's cosmology. The kSZ profile we presented in this paper corrects for this effect, by rescaling the kSZ profile by the ratio of  $\langle v_{\rm LOS}^2 \rangle^{\frac{1}{2}}$ measured in a simulation to $\langle v_{\rm LOS}^2 \rangle^{\frac{1}{2}}$ measured in fiducial \flamingo. Figure~\ref{fig:unscaled} shows the stacked kSZ profiles presented in Figure~\ref{fig:tksz_ggl_main}, but removing the correction for cosmic variance to the RMS of the radial velocities in the stack (Section~\ref{sec:cosmicvariance}), to illustrate its impact.

\section{The correlation between gas mass fraction and suppression of the matter power spectrum}

\begin{figure*}
\centering
\includegraphics[width=0.9\textwidth]{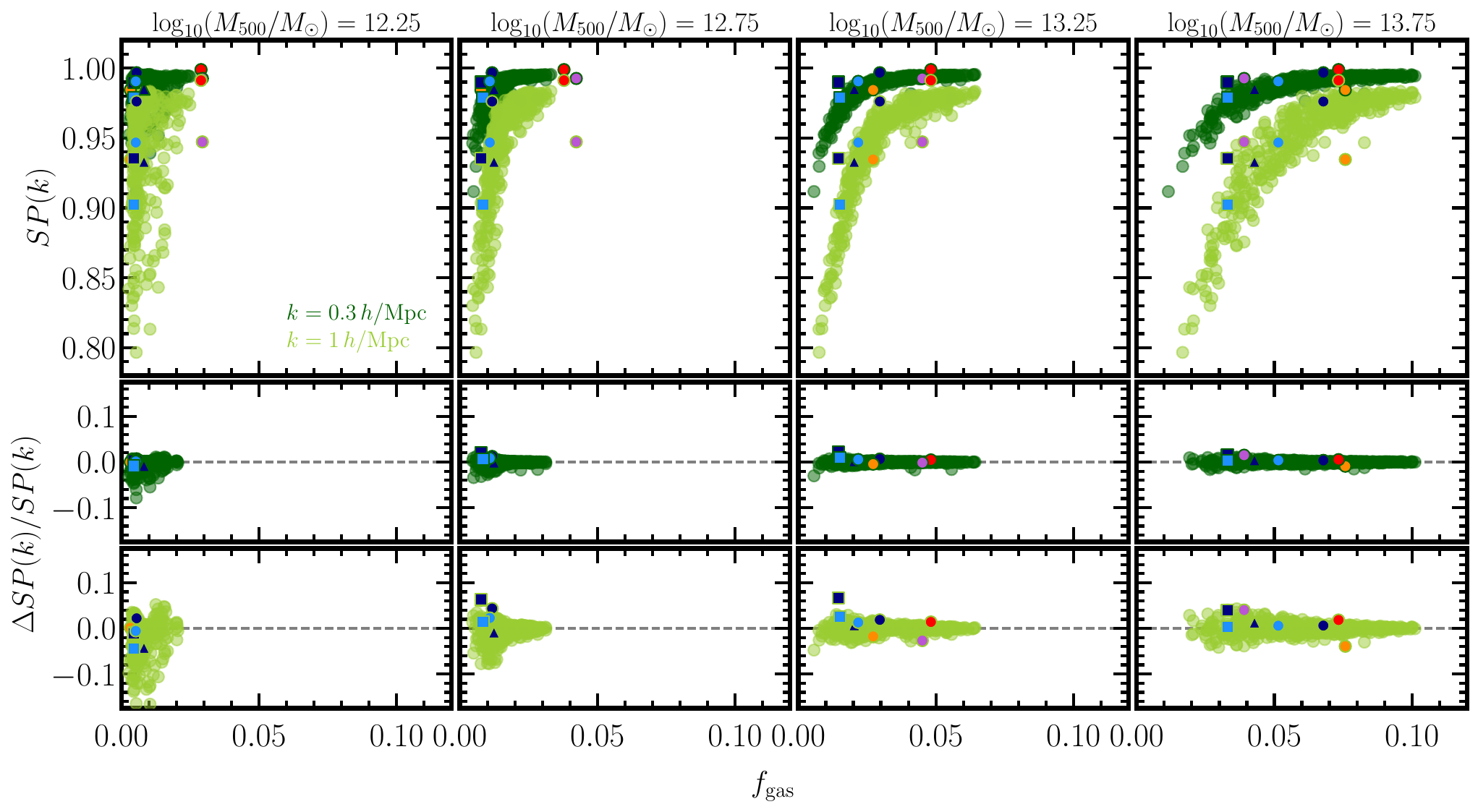}
\caption{As Figure~\ref{fig:antilles_tksz_corr}, but showing the relationship between the median gas mass fraction in halos, $f_{\mathrm{gas,500}}=M_{\mathrm{gas}}/M_{500}$ (where $M_{\mathrm{gas,500}}$ is the gas mass measured within $r_{500}$), and the suppression of the matter power spectrum, $SP(k)$.}
    \label{fig:fgas=pk}
\end{figure*}

Figure~\ref{fig:fgas=pk} shows the relationship between the ratio of gas mass to total mass measured within $r_{500}$ in groups and clusters, and the suppression of the matter power spectrum due to baryonic feedback.

\bsp	
\label{lastpage}
\end{document}